\documentclass[12pt]{JHEP3}
\usepackage{amssymb,epsf,graphicx}

\title{Wedge states in string field theory}
\author{Martin Schnabl \\ Center for Theoretical Physics, Massachusetts Institute of Technology,\\
Cambridge, MA 02139, USA  \\E-mail: \email{schnabl@lns.mit.edu}} 

\abstract{
The wedge states form an important subalgebra in the string field theory.
We review and further investigate their various properties. 
We find in particular a novel expression for the wedge states, 
which allows to understand their star products purely algebraically.
The method allows also for treating the matter and ghost sectors 
separately. It turns out, that wedge states with different matter and 
ghost parts violate the associativity of the algebra. 
We introduce and study also wedge states with insertions of local operators and
show how they are useful for obtaining exact results about convergence of level
truncation calculations. These results help to clarify the issue of anomalies
related to the identity and some exterior derivations in the string field algebra.
}
\preprint{MIT-CTP-3248 \\
MIT-LNS-02-296\\
{\tt hep-th/0201095}
}

\newcommand{\re}{\mathop{\rm Re}\nolimits}

\newcommand{\bpz}{\mathop{\rm bpz}\nolimits}

\newcommand{\ad}{\mathop{\rm ad}\nolimits}
\newcommand{\logit}{\mathop{\rm logit}\nolimits}

\def\bra#1{\langle #1 |}
\def\ket#1{|#1 \rangle}
\def\aver#1{\langle\, #1 \,\rangle}

\def\l{\left}
\def\r{\right}

\def \be {\begin{equation}}
\def \ee {\end{equation}}
\def \bea {\begin{eqnarray}}
\def \eea {\end{eqnarray}}
\def \bdm {\begin{displaymath}}
\def \edm {\end{displaymath}}

\def \nn {{\mathbb N}}
\def \zz {{\mathbb Z}}
\def \cc {{\mathbb C}}
\def \rr {{\mathbb R}}

\def \dd {{\cal D}}

\def \aa {{\cal A}}

\def \pp {{\cal P}}

\def \ccc {{\cal C}}

\def \sf  {string field }
\def \sft {string field theory }

\input{psfig}
\begin{document}
\section{Introduction}

Open string field theory \cite{WittenSFT,GJ1,GJ2,LPP1,LPP2} in the last two years
has experienced great renaissance as it turned out to be a
powerful tool for understanding nonperturbative off-shell
phenomenon of tachyon condensation in string theory. The famous
Sen's conjectures \cite{Sen:Descent,Sen:Universality}, by now has
been confirmed within the \sft to a high level of confidence
following the works \cite{SZ,MT}. For a complete lists of references
the reader is referred to the reviews
\cite{Ohmori,RSZ5,DeSmet,SchnablPhD,Arefeva}. Unfortunately, most of the checks
in the Witten's cubic \sft up to date were performed only
numerically. Notable exception is the idea of the vacuum \sft
(VSFT) \cite{RSZ1,GRSZ}, which is the standard \sft expanded around
the true nonperturbative vacuum. This approach seems quite promising,
it has already been possible to obtain some analytic results as for 
example the ratios of D-brane tensions \cite{RSZ4}.

The basic assumption of VSFT is that after suitable reparametrization of the 
string field,  the kinetic operator can be expressed entirely using the
ghost oscillators. The classical solutions corresponding to D-branes 
can then be found in terms of projectors in the matter sector of the 
\sf algebra. The only explicitly known (nontrivial) projector until 
very recently, has been the sliver state, which belongs to the family 
of wedge states, found by Rastelli and Zwiebach \cite{RZ}.
These states form a commutative subalgebra within the \sf algebra. 
Among other states in this family, there is also the $SL(2,\rr)$ 
invariant vacuum $\ket{0}$ and the
identity element $\ket{I}$ of the algebra.   

In this paper we would like to study various aspects of this important 
family of wedge states, expressing our results mainly in the so called 
universal basis formed by matter Virasoro operators and 
ghosts acting on the vacuum. We shall find a new expression for all the 
wedge states which enables us to prove algebraically the star
multiplication rules for the wedge states. We can also express all the
wedge states as explicit reparametrizations of the vacuum, which 
completes the geometrical picture of \cite{GRSZ}. The sliver and the
identity emerge as infinite reparametrizations of the vacuum and this
property makes them projectors. The methods developed allow us to
study so called unbalanced wedge states, which are factorized states
whose matter and ghost parts correspond to different wedge states.
We find that such states necessarily violate the cyclicity of the
three vertex and therefore also the associativity. The same happens
for wedge states in the twisted conformal field theory (CFT) of 
\cite{GRSZ} whose overall central charge is nonvanishing.

Another topic we discuss thoroughly is star multiplication of wedge 
states with insertions of local operators. The results can be used to
show, that in certain cases the level truncation actually breaks down. 
Second application of the rules for wedge states with insertions is to
study the anomalies related to the \sf algebra identity and some
exterior derivations.  We show in particular, that the object 
$c_0 \ket{I}$ should be better excluded from 
the star algebra, since it has ambiguous star products with the wedge 
states. As a byproduct we find a new `sum rule' for the tachyon
condensate in ordinary cubic \sf theory.

The paper is organized as follows. In section 2 we review some
basic facts about finite conformal transformation and wedge states.
In section 3 we introduce wedge states with insertions and derive 
their star-multiplication properties. The results are used to 
demonstrate that level expansion breaks down in the product of just 
four Fock space states and that this could possibly violate 
associativity. Section 4 deals with the behavior of
the coefficients which appear in the original definition \cite{RZ} 
of the wedge states. It introduces the mathematical concept of 
the iterative logarithm, which helps in section 5 to find a novel 
explicit formula for all the wedge states. This can be also rewritten
in a form which makes manifest that the wedge states are
reparametrizations of the star product. Particular attention is also
paid to matter and ghost sectors separately. We show that the wedge
states in these sectors do not have finite norm and generically lead
to violation of associativity. 
Finally section 6 is devoted to the star algebra identity. We show
explicitly that the object $c_0 \ket{I}$ has ambiguous star products 
with other states in the algebra.
Appendix A illustrates how the multiplication of wedge states behaves 
in the level truncation approximation and appendix B contains some graphs
related to the discussion in section 4.

\section{Wedge states}
\label{wedgestates}

\subsection{Finite conformal transformation}

Let us recall first some basic facts about finite conformal
transformations. A primary field $\Psi(z)$ of conformal weight $d$
transforms under finite conformal transformation $f$ as
\be
f \circ \Psi = \l[f'(z)\r]^d \Psi(f(z)).
\ee
We would like to rewrite this transformation rule using the
Virasoro generators $L_n$ of the conformal group in the form
\be
\l[f'(z)\r]^d \Psi(f(z)) = U_f \Psi(z) U_f^{-1},
\ee
where
\be
U_f = e^{ \sum v_n L_n}.
\ee
To determine the coefficients $v_n$ we note that
\be
U_f \Psi(z) U_f^{-1}= e^{\ad_{\sum v_n L_n}} \Psi(z),
\ee
where as usually $\ad_X Y = [X,Y]$. We may prove an important
identity
\be
\l(\ad_{\sum v_n L_n} \r)^k \Psi(z) = (v(z) \partial_z + d
v'(z))^k \Psi(z),
\ee
for any $k \in \nn$, where we set
\be
v(z) = \sum v_n z^{n+1}.
\ee
The proof for $k=1$ can be easily performed for example by
expanding $\Psi=\sum \frac{\Psi_n}{z^{n+d}}$ and using the
commutation relations
\be
[L_m,\Psi_n]=((d-1)m-n) \Psi_{m+n}.
\ee
For $k>1$ it then extends trivially. We thus see that in general
\be
U_f \Psi(z) U_f^{-1} = e^{v(z) \partial_z + d v'(z)} \Psi(z).
\ee
Our task is now for a given $f(z)$ to find a solution $v(z)$, such
that for any $\Psi(z)$ of an arbitrary dimension $d$ holds
\be
e^{v(z) \partial_z + d v'(z)} \Psi(z) = \l[f'(z)\r]^d \Psi(f(z)).
\ee
A priori, it is not even clear that such $v(z)$ exists. Let us
insert into the left hand side the identity $e^{-v\partial}
e^{v\partial}$. Since
\be
e^{v(z) \partial_z} \Psi(z) = \Psi(e^{v(z) \partial_z} z),
\ee
as one can easily check, and $e^{v(z) \partial_z + d v'(z)}
e^{-v(z)\partial_z}$ is just an ordinary function, we have to take
$v(z)$ such that
\be\label{defv}
e^{v(z) \partial_z} z = f(z).
\ee
From that follows another important relation
\be\label{Julia}
v(z) \partial_z f(z) = v(f(z)),
\ee
which in the mathematical literature is called the Julia equation.
The proof is simple:
\be
v(z) \partial_z f(z) = v(z) \partial_z e^{v(z) \partial_z} z =
e^{v(z) \partial_z} v(z) = v(f(z)).
\ee
For completeness and consistency we should be able to show also
\be
e^{v(z) \partial_z + d v'(z)} e^{-v(z) \partial_z} =
\l[f'(z)\r]^d,
\ee
for any $d$. In order to prove it let us define for $t \in [0,1]$
\bea
f_t(z) &=& e^{tv(z) \partial_z} z,
\nonumber\\
X_t (z) &=& e^{tv(z)\partial_z + dtv'(z)} e^{-tv(z) \partial_z},
\eea
and derive a differential equation for $X_t$:
\bea
\partial_t X_t(z) &=& d v'(f_t(z)) X_t(z)
\nonumber\\
&=& d \frac{\partial_t
\partial_z f_t(z)}{\partial_z f_t(z)} X_t(z).
\eea
Integrating this equation from $0$ to $t$ we obtain
\be
X_t(z)=  \l[f_t'(z)\r]^d,
\ee
which for $t=1$ completes our proof.

For a given analytic $f(z)$ we can formally determine $v(z)$ from
(\ref{defv}). Plugging the Laurent expansion of $f$ we get all the
coefficients $v_n$ recursively. If $f$ vanishes at the origin
$f(0)=0$ and is holomorphic nearby, then  only $v_n$ with $n \ge
0$ are nonzero.

An important property of the operators $U_f$, which follows
directly from their definition is
\be\label{UUcomp}
U_{f\circ g}=U_f U_g
\ee
for any two functions $f$ and $g$ holomorphic at the origin and obeying
$f(0)=g(0)=0$.\footnote{
This condition guarantees, that $f(g(z))$ is holomorphic in a finite neighborhood
of the origin. If for example $f(g(z))$ were holomorphic only in some annular region around 
zero, then there would be a c-number multiplicative anomaly in (\ref{UUcomp}) given by $e^{\kappa c}$, 
where $\kappa$ is a constant depending on the maps $f$ and $g$, and $c$ is the central charge of
the Virasoro algebra.} 
For some purposes it is convenient to separate out the global scaling
component $v_0$. This is easily achieved by writing
\be
f(z)=f'(0) \frac{f(z)}{f'(0)}
\ee
and using the composition rule (\ref{UUcomp}). It follows that
\be
U_f = e^{v_0 L_0} e^{ \sum_{n \ge 1} v_n L_n},
\ee
where
\bea\label{defvprime}
e^{v_0} &=& f'(0),
\nonumber\\
e^{\sum_{n \ge 1} v_n z^{n+1} \partial_z} &=& \frac{f(z)}{f'(0)}.
\eea

\subsection{The definition of wedge states}

The wedge states form a subset of more general surface states.
Surface state corresponding to a given conformal map $f(z)$ is
defined by
\be\label{surfacedef}
\aver{f | \phi} = \aver{f\circ \phi}, \qquad \forall \phi
\ee
and therefore
\be
\bra{f} = \bra{0} U_{f}.
\ee
Wedge states are defined as a one parameter family of surface
states associated to conformal mappings
\be
f_r(z)=h^{-1}\left(h(z)^{\frac{2}{r}}\right) =  \tan\left(
\frac{2}{r} \arctan(z) \right),
\ee
where $h(z)=\frac{1+iz}{1-iz}$. The mapping $f_r(z)$ first maps
half-disk in the upper half plane into a half-disk in the right
half-plane, then shrinks or expands it into a wedge of angle
$\frac{360^\circ}{r}$ and finally maps it back into the upper
half-plane. We shall denote the wedge states as
\be
\bra{r} \equiv \bra{f_r} = \bra{0} U_{f_r}.
\ee
The associated kets are
\be
\ket{r} = U_r^\dagger \ket{0},
\ee
where $U_r^\dagger$ is defined as BPZ conjugation, which means
here that $L_n$ gets replaced by $(-1)^n L_{-n}$.

The one point function (\ref{surfacedef}) on the upper half plane
can be alternatively calculated on the disk via
\be\label{diskcorr}
\aver{f_r \circ \phi}_ {\rm half-plane} = \aver{F_r \circ
\phi}_{\rm disk},
\ee
where $F_r(z) = h(z)^{\frac{2}{r}}$. From the results of
\cite{LPP1} we know that we can apply any conformal
transformation, not necessarily $SL(2,\cc)$, to any correlator.
Since only $SL(2,\cc)$ transformations map the complex plane into
itself in a single valued manner, a general mapping $f(z)$ will
carry the plane into a Riemann surface with branch points.
Evaluation of conformal field theory correlators on a general
Riemann surface has to be defined, the most natural choice is to
evaluate the propagators $\aver{XX}$ and $\aver{bc}$ by mapping
them back to the plane. It would seem that we have not gained
anything, the bonus comes later when we glue together various
pieces of Riemann surfaces.

By simple mapping $z \to z^{\frac{r}{2}}$ the disk correlator can
be viewed as ordinary one point function on the Riemann surface
with total opening angle $\pi r$.

\FIGURE{\begin{picture}(0,0)%
\includegraphics{halfdisk.pstex}%
\end{picture}%
\setlength{\unitlength}{3947sp}%
\begingroup\makeatletter\ifx\SetFigFont\undefined%
\gdef\SetFigFont#1#2#3#4#5{%
  \reset@font\fontsize{#1}{#2pt}%
  \fontfamily{#3}\fontseries{#4}\fontshape{#5}%
  \selectfont}%
\fi\endgroup%
\begin{picture}(5085,1395)(1504,-4036)
\put(3505,-3339){\makebox(0,0)[lb]{\smash{\SetFigFont{7}{8.4}{\rmdefault}{\mddefault}{\updefault}
$h \circ \phi (0)$
}}}
\put(6027,-3340){\makebox(0,0)[lb]{\smash{\SetFigFont{7}{8.4}{\familydefault}{\mddefault}{\updefault}
$\pi$
}}}
\put(1504,-3331){\makebox(0,0)[lb]{\smash{\SetFigFont{7}{8.4}{\rmdefault}{\mddefault}{\updefault}
$h \circ {\cal V}_{\ket{r}}$
}}}
\put(5148,-3340){\makebox(0,0)[lb]{\smash{\SetFigFont{7}{8.4}{\familydefault}{\mddefault}{\updefault}
$\pi(r\!-\!1)$
}}}
\put(6589,-3339){\makebox(0,0)[lb]{\smash{\SetFigFont{7}{8.4}{\rmdefault}{\mddefault}{\updefault}
$h \circ \phi (0)$
}}}
\end{picture}

\caption{\small Graphical representation of the BPZ inner product of the
wedge state $\ket{r}$  and one auxiliary state $\ket{\phi}$. This can
be calculated as a one point function on the Riemann surface with total
opening angle $\pi r$.}
\label{Fig2vertex}
}

BPZ contraction in (\ref{surfacedef}) can in general be viewed as
two point function on the disk, where at point $1$ we insert the
vertex operator creating the state $\phi$ and in $-1$ the vertex
operator for the wedge state. The functional integral over the
left and right half-disk  separately with fixed boundary condition
on the line segment separating them, produces two Schr\"odinger
functionals for these two states. The functional integral over the
boundary between the half-disks represents the BPZ contraction
itself.

From all of this discussion it should be clear that gluing in
half-disk with insertion of the vertex operator for the wedge
state (which we do not know explicitly) is equivalent to gluing a
piece of Riemann surface of total opening angle $\pi(r-1)$.

The star multiplication of wedge states readily follows
(see \cite{RZ,RSZ4,David}).
The three vertex contracted with two wedge states $\ket{r}, \ket{s}$
and one auxiliary state $\ket{\phi}$
\be
\bra{V} \ket{r} \otimes \ket{s} \otimes \ket{\phi}
\ee
depicted at Fig. \ref{Fig3vertex} can be represented first as Riemann surface of 
total opening angle $3\pi$ with three insertions.
\FIGURE{
\begin{picture}(0,0)%
\includegraphics{trisect.pstex}%
\end{picture}%
\setlength{\unitlength}{3947sp}%
\begingroup\makeatletter\ifx\SetFigFont\undefined%
\gdef\SetFigFont#1#2#3#4#5{%
  \reset@font\fontsize{#1}{#2pt}%
  \fontfamily{#3}\fontseries{#4}\fontshape{#5}%
  \selectfont}%
\fi\endgroup%
\begin{picture}(7489,1455)(1426,-4068)
\put(3247,-2697){\makebox(0,0)[lb]{\smash{\SetFigFont{7}{8.4}{\rmdefault}{\mddefault}{\updefault}
$h^{\frac{2}{3}} \circ \phi (0)$
}}}
\put(3247,-4068){\makebox(0,0)[lb]{\smash{\SetFigFont{7}{8.4}{\rmdefault}{\mddefault}{\updefault}
$h^{\frac{2}{3}} \circ {\cal V}_{\ket{r}}$
}}}
\put(8389,-3168){\makebox(0,0)[lb]{\smash{\SetFigFont{7}{8.4}{\familydefault}{\mddefault}{\updefault}
$\pi$
}}}
\put(8732,-2697){\makebox(0,0)[lb]{\smash{\SetFigFont{7}{8.4}{\rmdefault}{\mddefault}{\updefault}
$h \circ \phi (0)$
}}}
\put(5647,-3168){\makebox(0,0)[lb]{\smash{\SetFigFont{7}{8.4}{\familydefault}{\mddefault}{\updefault}
$\pi$
}}}
\put(5561,-3597){\makebox(0,0)[lb]{\smash{\SetFigFont{7}{8.4}{\familydefault}{\mddefault}{\updefault}
$\pi$
}}}
\put(5989,-2697){\makebox(0,0)[lb]{\smash{\SetFigFont{7}{8.4}{\rmdefault}{\mddefault}{\updefault}
$h \circ \phi (0)$
}}}
\put(5989,-4068){\makebox(0,0)[lb]{\smash{\SetFigFont{7}{8.4}{\rmdefault}{\mddefault}{\updefault}
$h \circ {\cal V}_{\ket{r}}$
}}}
\put(5132,-3340){\makebox(0,0)[lb]{\smash{\SetFigFont{7}{8.4}{\familydefault}{\mddefault}{\updefault}
$\pi$
}}}
\put(4276,-3331){\makebox(0,0)[lb]{\smash{\SetFigFont{7}{8.4}{\rmdefault}{\mddefault}{\updefault}
$h \circ {\cal V}_{\ket{s}}$
}}}
\put(1426,-3331){\makebox(0,0)[lb]{\smash{\SetFigFont{7}{8.4}{\rmdefault}{\mddefault}{\updefault}
$h^{\frac{2}{3}} \circ {\cal V}_{\ket{s}}$
}}}
\put(7576,-3340){\makebox(0,0)[lb]{\smash{\SetFigFont{7}{8.4}{\familydefault}{\mddefault}{\updefault}
\mbox{$\pi(s\!-\!1)$}
}}}
\put(8303,-3597){\makebox(0,0)[lb]{\smash{\SetFigFont{7}{8.4}{\familydefault}{\mddefault}{\updefault}
$\pi(r\!-\!1)$
}}}
\end{picture}

\caption{\small Graphical representation of the 3-vertex
contracted with two wedge states $\ket{r}$, $\ket{s}$ and one
auxiliary state $\ket{\phi}$. By a conformal mapping we map the
disk to a Riemann surface with opening angle $3\pi$ and then
replace  two of the half-disks by helixes of angles $\pi(r-1)$ and
$\pi(s-1)$ respectively. } \label{Fig3vertex}
}
By the above mentioned equivalence
we can replace the half-disks with vertex operators for the wedge
states by parts of the Riemann surfaces of angles $\pi(r-1)$ and
$\pi(s-1)$. Gluing them together produces a surface with total
angle $\pi(r+s-2)$. Equating $r+s-2 = t-1$ gives $t=r+s-1$ and
thus the desired composition rule
\be\label{wedgecomp}
\ket{r} * \ket{s} = \ket{r+s-1}.
\ee
Let us give some concrete examples of the wedge states. Using the
recursive relations following from (\ref{defvprime}) we have
\bea\label{wedgeex}
\ket{1} &=& e^{L_{-2} - \frac12 L_{-4} + \frac12 L_{-6} - \frac{7}{12} L_{-8}
+ \frac{2}{3} L_{-10} - \frac{13}{20} L_{-12}+ \cdots}\ket{0}
\nonumber\\
\ket{2} &=& \ket{0}
\nonumber\\
\ket{3} &=& e^{-\frac{5}{27} L_{-2} +\frac{13}{486} L_{-4} -
\frac{317}{39366} L_{-6} + \frac{715}{236196} L_{-8}
- \frac{17870}{14348907} L_{-10} + \cdots} \ket{0}
\nonumber\\
\ket{4} &=& e^{-\frac{1}{4} L_{-2} +\frac{1}{32} L_{-4} -
\frac{1}{128} L_{-6} + \frac{7}{3072} L_{-8}
- \frac{1}{1536} L_{-10} + \cdots}\ket{0}
\nonumber\\
\ket{\infty} &=& e^{-\frac{1}{3} L_{-2} +\frac{1}{30} L_{-4} -
\frac{11}{1890} L_{-6} + \frac{1}{1260} L_{-8}
+\frac{34}{467775} L_{-10} + \cdots}\ket{0}.
\eea
For general $r$ the wedge state looks as 
\bea\label{wedger}
\ket{r} &=& \exp\l(
-\frac{r^2 -4}{3\,r^2} \,L_{-2}
+\frac{r^4-16}{30\,r^4}\, L_{-4}
-\frac{\l(r^2-4\r) \, \l( 176 + 128\,r^2 + 11\,r^4\r)}{1890\,r^6} \, L_{-6}+
\r.
\nonumber\\
&& \l.
\qquad +\frac{\l(r^2-4 \r) \,\l( r^2 +4\r) \,
     \l( 16 + 32\,r^2 + r^4 \r) }{1260\,r^8} \, L_{-8}
+ \cdots \r) \ket{0}.
\eea
To avoid confusion, the state $\ket{0}$ always denotes the
$SL(2,\rr)$ invariant vacuum, which is a wedge state $\ket{2}$.
Wedge state with $r=0$ simply does not exist.
The state $\ket{1}$ is the identity of the star algebra and will be
discussed further in section \ref{Identity}. Various aspects of the
identity has already been studied in 
\cite{GJ1,GJ2,Horowitz,QS,HM,Romans,RZ,EFHM,Matsuo2,TT,Kishimoto,KO}.
The limiting state $\ket{\infty}$, known as a sliver, is a projector in
the star algebra. 
Since its matter part provides us with a solution to the vacuum \sf
theory, it has been thoroughly studied in the literature
\cite{RZ,KP,RSZ2,RSZ3,RSZ4,GT,Matsuo1,Mukhopadhyay,Moeller,Moore,AGM,MS,Bonora}.

\section{Wedge states with insertions}
\label{WedgeInsert}
\subsection{Basic properties}

Let us take  a primary field $\pp (z)$ of dimension $d$ and a point $x$
inside the unit circle.\footnote{One may try to go outside of the unit circle by
an analytic continuation, but it is quite problematic. Our formulas show clearly that
for $x \to \pm i$ the level truncation breaks down, the star product itself is
singular. There are two branch cuts starting at $\pm i$ and going to infinity.
Across these branch cuts the star product would vary discontinuously and
therefore it would fail to be a good product.}
The wedge states with insertion are defined by
\be\label{defwwi}
\bra{f_{r,\pp,x}} = \bra{0} I \circ \pp (x) \, U_{f_r},
\ee
where $I z = - 1/z$. Written as kets they read
\be\label{defwwiket}
\ket{f_{r,\pp,x}} =  U_{f_r}^\dagger  \pp (x) \ket{0}.
\ee
More generally we can have any number of insertions. 
From the basic property of conformal field theory on Riemann surfaces
\bea
\aver{f_{r,\pp,x}|\phi} &=& \aver{ h \circ I \circ \pp (x) \,\,
h^{\frac{2}{r}} \circ \phi(0)}_{\rm disk}
\nonumber\\
&=& \aver{ h^{\frac{r}{2}} \circ I \circ \pp (x) \,\, h \circ
\phi(0)}_{\rm Riemann-surface}
\eea
we see that the effect of the vertex operator for the wedge
state with insertion is again to replace this half-disk with a
piece of Riemann surface of total opening angle $\pi(r-1)$ and
inserting an operator $\pp$ at point
\be
h^{\frac{r}{2}} \circ I (x) = e^{i r \arctan x + i\frac{\pi
r}{2}}.
\ee
This equality is actually valid for the standard definition of the function
$\arctan x $ in the complex plane. However to appreciate the geometric picture, it
is better to temporarily think of $x$ as sitting in the line segment $(-1,1)$ of
the real axis.  Let us now calculate the star product
\be
U_r^\dagger \pp_1(x) \ket{0} * U_s^\dagger \pp_2(y) \ket{0}.
\ee
Again we consider the Witten vertex as Riemann surface obtained by
gluing three half-disks, corresponding to the states $U_r^\dagger
\pp_1(x) \ket{0},  U_s^\dagger \pp_2(y) \ket{0}$ and $\ket{\phi}$
in clockwise order. We can replace two of them according to the
above rule. Finally we wish to reinterpret this three vertex as a
BPZ contraction of $\phi$ and a wedge state with two insertions.
To find the insertion points we have to match simply
\bea
e^{i s \arctan y + i\frac{\pi s}{2}} &=& e^{i t \arctan y' +
i\frac{\pi t}{2}},
\nonumber\\
e^{i r \arctan x + i\frac{\pi r}{2} + i \pi (s-1)} &=& e^{i t
\arctan x' + i\frac{\pi t}{2}}
\eea
where $t=r+s-1$. The solution is simply
\bea
x' \equiv g_1(x) &=& \cot \l( \frac{r}{t} \l(+\frac{\pi}{2} - \arctan x\r) \r),
\nonumber\\
y' \equiv g_2(y) &=& \cot \l( \frac{s}{t} \l(-\frac{\pi}{2} - \arctan y\r) \r),
\eea
which is manifestly holomorphic in the whole unit disk. Alternatively,\footnote{
I thank I.~Ellwood for helpful conversations on this issue.} one can write 
these functions as
\bea
x' &=&  h^{-1}\l(e^{+i \pi(1-\frac{r}{t})} (h(x))^{\frac{r}{t}}\r),
\nonumber\\
y' &=&  h^{-1}\l(e^{-i \pi(1-\frac{s}{t})} (h(y))^{\frac{s}{t}}\r).
\eea

Having found out the insertion points it is quite simple to work out also the
normalization factors coming from the transformation law of the primary fields
$\pp_1$ and $\pp_2$. Altogether, we arrive at
\bea\label{wwi}
U_r^\dagger \pp_1(x) \ket{0} * U_s^\dagger \pp_2(y) \ket{0} &=&
\l(\frac{r}{t}\cdot\frac{1+x'^2}{1+x^2} \r)^{d_1}
\l(\frac{s}{t}\cdot\frac{1+y'^2}{1+y^2} \r)^{d_2}
U_{r+s-1}^\dagger \pp_1(x') \pp_2(y') \ket{0}
\nonumber\\\label{wwigen}
&=& U_{r+s-1}^\dagger \, g_1 \circ \pp_1(x) \, g_2 \circ \pp_2(y)
\ket{0}
\eea
Although we have derived the formula for primary fields, because
it was easy to trace the insertion points, the resulting formula
(\ref{wwigen}) is actually valid for all local fields and can be
used as long as we know the transformation properties of the
fields.

Let us check this formula on few examples. 
For the star products with ghost insertions
\bea
c(0) \ket{0} * c(0) \ket{0} &=& \l(\frac{9}{8}\r)^2 U_3^\dagger\, c\!\l(\frac{1}{\sqrt{3}}\r)
 c\!\l(-\frac{1}{\sqrt{3}}\r) \ket{0},
\nonumber\\
\ket{0} * c(0) \ket{0} &=& \frac{9}{8}\, U_3^\dagger\, 
 c\!\l(-\frac{1}{\sqrt{3}}\r) \ket{0},
\eea
we get the same results as obtained previously in \cite{RZ}. 
For the star products with energy momentum tensor insertions 
we find
\bea
\ket{0} * T(0) \ket{0} &=& \l(\frac{8}{9}\r)^2 U_3^\dagger\, 
 T\!\l(-\frac{1}{\sqrt{3}}\r) \ket{0},
\nonumber\\
T(0) \ket{0} * \ket{0} &=& \l(\frac{8}{9}\r)^2 U_3^\dagger\, 
 T\!\l(\frac{1}{\sqrt{3}}\r) \ket{0},
\nonumber\\
T(0) \ket{0} * T(0) \ket{0} &=& \l(\frac{8}{9}\r)^4 U_3^\dagger\, T\!\l(\frac{1}{\sqrt{3}}\r)
 T\!\l(-\frac{1}{\sqrt{3}}\r) \ket{0},
\nonumber\\
&=& \l(\frac{8}{9}\r)^4 U_3^\dagger\, e^{-\frac{1}{\sqrt{3}} L_{-1}} 
T\!\l(\frac{2}{\sqrt{3}}\r) T(0) \ket{0}.
\eea
These formulas can be tested exactly to any given level by explicit
calculation of the star products using for instance the conservation
laws of \cite{Samuel, GJ2, RZ}.
Less trivial example we have tested is 
\bea
L_{-2}\ket{0} * \ket{\infty} &=& \l(\frac{2}{\pi}\r)^4 {U'}_\infty^\dagger
T\l(\frac{2}{\pi}\r)\ket{0}
\\\nonumber
&\simeq& 0.164 L_{-2}\ket{0}+0.105 L_{-3}\ket{0}+0.067L_{-4}\ket{0}
-0.055L_{-2} L_{-2}\ket{0}+\cdots.
\eea
Here the prime on ${U'}_\infty^\dagger$ means that the divergent factor
$\l(\frac{2}{r}\r)^{L_0}$ has been omitted from its expression.
Numerical calculation at level 20 gives result
\be
L_{-2}\ket{0} * \ket{\infty} \simeq
0.180 L_{-2}\ket{0}+0.110 L_{-3}\ket{0}+0.067 L_{-4}\ket{0}
-0.058 L_{-2} L_{-2}\ket{0}+\cdots,
\ee
which is in a reasonable agreement.

\subsection{Breakdown of the level truncation}

Now we would like to use the formula (\ref{wwi}) to argue that the
level truncation calculation breaks down in quite simple cases. 
Imagine we wish to calculate
\be
L_{-2}\ket{0} * \underbrace{\ket{0} * \ket{0}
*\ket{0}*\cdots *\ket{0}}_{(r-1)\,\, \mbox{\small times}}. 
\ee 
Relying on the associativity of the star product, there are many ways
to do the calculation. The easiest possibility is to multiply first
all the vacua on the right to get wedge state $\ket{r}$ and then to
use our formula (\ref{wwi}) to find 
\be
L_{-2}\ket{0} * U_r^\dagger \ket{0} = \l[ \frac{r+1}{2} \sin^2
\frac{\pi}{r+1} \r]^{-2} U_{r+1}^\dagger T\l(\cot\frac{\pi}{r+1}\r)
\ket{0}.
\ee
For general primary fields of dimension $d$ the formula would look the
same with the exponent $-2$ replaced by $-d$. On the other hand we may
try to calculate the star product successively, as indicated by the brackets
\be
((((L_{-2}\ket{0} * \ket{0}) * \ket{0})
*\ket{0})*\cdots *\ket{0}). 
\ee 
This way we get 
\bea
L_{-2} \ket{0} * \ket{0} &=& \l(\frac{8}{9}\r)^2 U_3^\dagger\, 
 T\!\l(\frac{1}{\sqrt{3}}\r) \ket{0},
\nonumber\\
(L_{-2} \ket{0} * \ket{0})*\ket{0}  &=&  U_4^\dagger\, T(1) \ket{0},
\nonumber\\
((L_{-2} \ket{0} * \ket{0})*\ket{0})*\ket{0}  &\stackrel{?}{=}&  \l[\frac{5}{2} \sin^2 \frac{\pi}{5} \r]^{-2}
U_5^\dagger\, T\l(\cot\frac{\pi}{5}\r)
\ket{0}.
\eea
The right hand side in the last equation follows by straightforward
use of the formula (\ref{wwi}). The reason why we put the question
mark above the equality sign is that in reality
\be
U_4^\dagger\, T(1) \ket{0} *\ket{0} 
\ee
is divergent in the level expansion and cannot be calculated
unambiguously. To show that consider 
\be
U_r^\dagger\, T(z) \ket{0} *\ket{0} \approx 
\l[ \frac{r}{r+1} \, e^{i \frac{\pi}{2} \frac{1}{r+1}}   
\, 2^{-\frac{r}{2(r+1)}} \, (1+iz)^{-\frac{r+2}{2(r+1)}} \r]^2 
U_{r+1}^\dagger\, T(i) \ket{0}
\ee
for $z \to i$. This is clearly divergent in this limit and would be
also divergent for any other primary with positive dimension. It means 
that $U_r^\dagger\, T(z) \ket{0} *\ket{0}$ as a series in $z$ has 
radius of convergence equal to $1$. Therefore the star product 
$U_4^\dagger\, T(1) \ket{0} *\ket{0}$ is not absolutely convergent in
level truncation. The result depends on the order of summation
and is thus ambiguous. 

\subsection{An alternative definition}

When we defined the wedge states with insertions in (\ref{defwwiket})
the reader could have asked why we did not define them simply as 
\be\label{altdefwwiket}
\pp(x) \, U_r^\dagger  \ket{0}.
\ee
This expression as it stands (assuming for a moment that $r \ne 2$) 
is convergent in the level expansion only 
for $|x|\ge 1$. To see that, we have to normal order the operator acting on 
the vacuum to get rid of all positively moded operators annihilating 
the vacuum. We get 
\be\label{twowwirel}
\pp(x) \, U_r^\dagger  \ket{0} = U_r^\dagger \, I\circ f_r \circ I
\circ \pp(x) \ket{0}
\ee
and we see that in order for the left hand side to be well defined, 
the series expansion of $I\circ f_r \circ I (x)$ in $x$ has to be convergent.

Now it seems, that we could simply combine the formulas (\ref{wwigen}) and (\ref{twowwirel})
to find a star product of states like (\ref{altdefwwiket}).
We have to be little careful though. In order to use the formula 
(\ref{wwigen}) legitimately, one would need
$x \in \dd_r$,  $y \in \dd_s$, where 
\be
\dd_r = \l\{ z; \, |I \circ f_r \circ I (z)| <1 \r\}.
\ee 
For $r\ge 4$ one can check that $\dd_r = \emptyset$. For $2 \le r < 4$ the domain
$\dd_r$ is nontrivial, but its 
intersection with the exterior of the unit circle $|x| \ge 1$ is still empty.
Only in the region $1 \le r < 2$ the intersection is not empty.    

Working now in the appropriate region of parameters $x,y,r$ and $s$ we 
can combine equations (\ref{wwigen}) and (\ref{twowwirel}).  
Assuming further $\re x>0$ and $\re y<0$ the resulting formula 
considerably simplifies
\be\label{wwialt}
\pp(x) \, U_r^\dagger  \ket{0}  * \pp(y) \, U_s^\dagger  \ket{0}
= \pp(x) \pp(y) \, U_{r+s-1}^\dagger  \ket{0},
\ee
which actually remains true for arbitrary $r,s \ge 1$, $\re x>0$ and $\re y<0$ 
by an analytic continuation of the formulas (\ref{wwigen}) and (\ref{twowwirel}).
It is an obvious manifestation of the fact, that the left part of the first string
(region $\re x>0$) will become the left part of the product, the right part 
of the second string (region $\re y<0$) then becomes the right part of the product.

\section{Behavior of the wedge state coefficients}

Looking at the examples of wedge states (\ref{wedgeex}) or on the
general formula (\ref{wedger}) one may wonder whether there is
some closed expression for all the coefficients $v_n$. Another
question one may ask, is what is the behavior of $v_n$ for large
$n$. First impression is that for $r=1$ the coefficients somehow
chaotically oscillate between $\pm 1$, whereas for $r>2$ they
decrease exponentially to zero.

To obtain the expressions for the operators $U_f$ with the global
scaling component separated we need to solve the equation
(\ref{defv}) or (\ref{Julia}) with
\be
\tilde f_r(z)=\frac{r}{2} h^{-1}\left(h(z)^{\frac{2}{r}}\right) =
\frac{r}{2} \tan\left( \frac{2}{r} \arctan(z) \right)
\ee
which satisfies $\tilde f_r'(0)=1$.  Given a function $f$
holomorphic in the neighborhood of the origin one can always look
for analytic solutions $v(z)$ to the Julia equation (\ref{Julia})
in terms of formal power series (FPS). The solution is unique up
to an overall constant which can be fixed for $f$ of the form
\be\label{fseries}
f(z)=z+\sum_{n=m}^\infty b_n z^n,  \qquad   b_m \ne 0, m \ge 2
\ee
by requiring
\be
v(z) = b_m z^m + \sum_{n=m+1}^\infty c_n z^n.
\ee
Note that precisely with this normalization the function $v(z)$
satisfies also the (\ref{defv}). Such a unique solution is called
the {\it iterative logarithm} and denoted either as $f_*$ or
$\logit f$. Interesting problem is when this FPS has finite radius
of convergence.

It has been proved that if $f$ is a meromorphic function, regular
at the origin and having the expansion (\ref{fseries}) then the
formal power series $f_*$ has a positive radius of convergence if
and only if
\be
f(z)=\frac{z}{1+bz}, \qquad b \in \cc
\ee
This theorem is implied by the results of I.N.~Baker, P.~Erd\H{o}s
and E.~Jabotinsky, see \cite{Kuczma}.

Let us see how this result applies to our wedge states. All of the
functions $\tilde f_r$ are holomorphic near the origin, but only
those with $r=\frac 2k, \, k \in \zz$ are meromorphic in the whole
complex plane. Apart from the vacuum state $\ket{2}$ all the other
$\ket{1}, \ket{\frac{2}{3}}, \ldots$ thus correspond to divergent
FPS with zero radius of convergence.

What about the other wedge states? Using the Julia equation and
checking the overall normalization one can establish following
general properties of the iterative logarithm:
\bea
\logit f &=& - \logit f^{-1},
\\\label{logit2nd}
\logit (\phi^{-1} \circ f \circ \phi) &=& \frac{1}{\phi'} \l(
(\logit f) \circ \phi \r),
\eea
where $\phi(z)$ is any analytic function with $\phi(0)=0$ and
$\phi'(0) \neq 0$. From these two relations follows (by taking
$\phi(z) = \frac r2 z$)
\be
\logit \tilde f_{\frac{4}{r}} = -\frac{2}{r} \circ (\logit \tilde
f_r) \circ \frac{r}{2}.
\ee
We thus obtain for the Laurent coefficients of $v^{(r)}=\logit
\tilde f_r$ important relation
\be
v_k^{\l(\frac{4}{r}\r)} = - \l(\frac{2}{r}\r)^k v_k^{(r)},
\ee
which can be readily checked for the explicit expression
(\ref{wedger}). We see that the FPS $\logit \tilde f_r$ and
$\logit \tilde f_{\frac{4}{r}}$ have both either zero or finite
radius of convergence simultaneously. Summarizing, the FPS
corresponding to the vector field $v(z)$ has zero radius of
convergence for all $r=\frac 2k$ and $r=2k$ for $k\in\zz, k>1$. By
a limiting procedure this applies in particular to the interesting
sliver state $\ket{\infty}$. For $r=2k+1$ we cannot 
simply extend the above argument. One would have to generalize 
the above quoted mathematical theorem or perhaps rather choose
a different method. Nevertheless we expect qualitatively the same 
kind of behavior.   

The absence of any finite radius of convergence means that
starting from a certain level, some of the coefficients $v_n$
start to grow faster than any exponential. This rather surprising
result is confirmed by the actual calculation of the coefficients
up to $v_{100}$ which we have plotted for several wedge states in
the Appendix \ref{WedgeCoefficients}. All the coefficients were
calculated exactly using the recursive formula following from
(\ref{defv}).

To summarize we have shown that the Laurent expansion of $v(z)$
has zero radius of convergence and therefore the function $v(z)$
has an essential singularity at zero. This can be verified
independently by studying its poles or zeros. Using the Julia
equation one can find an infinite sequence of poles or zeros
approaching the origin. Its existence proves that there is an essential
singularity. This infinite sequence is easily found numerically, 
to prove analytically that it is actually infinite seems tedious. 

We suspect that the uncontrollable growth of the wedge state
coefficients is just an artifact of the symmetrical ordering of
the Virasoro generators. It seems likely that for other choices of
ordering the coefficients behave much better as was shown for the
identity $\ket{I}$ in \cite{EFHM}.

If this were not the case, then the series itself could be trusted
as asymptotic only. The success of level truncation for the star
products of wedge states (see Appendix \ref{AppStarProds}) would
appear to be analogous to the situation in QED, where at first few
orders the perturbation theory works perfectly well, but at higher
orders it breaks down. From the graphs in the Appendix
\ref{WedgeCoefficients} one can see that for the coefficients up
to about $v_{20}$ the coefficients decrease exponentially, this is
the basic reason why the low order calculations work well.

Finally let us comment on one technical aspect of the calculation.
To calculate e.g. the 100-th derivative at zero of the function
$\tilde f_r$ for generic $r$ directly, is beyond the capacity of
any computer. This problem however can be overcome by the
following formula for $n$ odd
\be
\frac{d^n}{dx^n} \tan\l(\frac 2r \arctan x \r)|_{x=0}=
\sum_{k=1,3,5,\ldots,n}  \l(\frac{2}{r}\r)^k
\frac{2^{k+1}(2^{k+1}-1)}{(k+1)!} B_{k+1} F(n,k),
\ee
where
\be
F(n,k)=\sum_{m_i; \sum m = \frac{n-k}{2}}
\frac{1}{(2m_1+1)\ldots(2m_k+1)}
\ee
is easily calculable recursively and $B_n$'s are the Bernoulli
numbers. For $n$ even the derivative is obviously zero.

\section{Wedge states and reparametrizations}

We start by deriving new expression for the wedge states which is
useful for many explicit analytic calculations. Given the finite
conformal transformation
\be
f(z)=\tan\l(\frac{2}{r}\arctan z\r)
\ee
we can easily find the vector field $v(z)$ which generates it.
From equation (\ref{logit2nd}) we get
\bea\label{defvnew}
v(z) &=& \logit f(z) = \logit \l[\arctan^{-1} \circ \frac{2}{r}
\arctan (z) \r]=
\nonumber\\
&=& \log\frac{2}{r} \cdot (1+z^2) \arctan z.
\eea
The associated finite conformal transformation operator is
\bea\label{Udef}
U_r &=& e^{\oint\!\frac{dz}{2\pi i}\, v(z) T(z) }
\nonumber\\
&=& e^{2\log\frac{r}{2} \l( -\frac 12 L_0 + \sum_{k=1}^\infty
\frac{(-1)^k}{(2k-1)(2k+1)} L_{2k} \r)}.
\eea
This is a nice expression since it makes manifest the properties
(\ref{UUcomp})
\bea\label{comprule}
U_r U_s &=& U_{\frac{rs}{2}},
\nonumber\\
U_2 &=& 1.
\eea
The wedge states are given by
\be
\ket{r} = U_r^\dagger \ket{0} = e^{2\log\frac{r}{2} A^\dagger}
\ket{0},
\ee
where we denote
\be
A= -\frac 12 L_0 +
\sum_{k=1}^\infty \frac{(-1)^k}{(2k-1)(2k+1)} L_{2k}
\ee
and $A^\dagger$ is its BPZ conjugate.\footnote{While this paper
was being written, this vector field (without the $L_0$ part)
appeared in a related context in \cite{RSZ7}.}  Calculating the
commutator
\be\label{K1Acomm}
[K_1,A] = \frac 12 K_1
\ee
we see that all wedge states manifestly obey
\be\label{K1conserv}
K_1 \ket{r} = 0,
\ee
a conservation law first found in \cite{RSZ2} and recently
discussed in \cite{RSZ7}. Next we can calculate the commutator
\be
[A,A^\dagger] = -\frac{1}{2} (A+A^\dagger) + \frac{c}{12}
\sum_{k=1}^\infty \frac{2k}{(2k-1)(2k+1)},
\ee
keeping the central charge $c$ for the time being arbitrary. The
last term in the commutator is logarithmically divergent, we could
set $c=0$, but it is interesting to continue with nonzero $c$
assuming some convenient regularization. Now it is useful to
introduce a new operator $B=-2A$ with the obvious commutation
relation
\be
[B,B^\dagger] = B + B^\dagger + \lambda,
\ee
where
\be
\lambda =  \frac{c}{3} \sum_{k=1}^\infty \frac{2k}{(2k-1)(2k+1)}.
\ee
Then one can easily find
\be
B {B^\dagger}^n \ket{0} = (B^\dagger +\lambda) \l[ \l(B^\dagger+1\r)^n -
{B^\dagger}^n \r] \ket{0}
\ee
and
\be\label{BeBdagger}
B\, e^{\beta B^\dagger} \ket{0} =  \l(e^\beta -1\r) (B^\dagger +
\lambda) \, e^{\beta B^\dagger} \ket{0}.
\ee
We are interested in particular in
\be
X(\alpha,\beta) = e^{\alpha B} e^{\beta B^\dagger} \ket{0},
\ee
which can be found by solving the differential equation
\be
\l(\partial_\alpha - \l(e^\beta -1\r)(\partial_\beta +\lambda) \r)
X(\alpha,\beta) = 0.
\ee
with the obvious initial condition $X(0,\beta)=e^{\beta B^\dagger}
\ket{0}$. The easiest way is to solve first the equation for
$\lambda=0$ and then to recover the correct $\lambda$ dependence
by replacing formally $B^\dagger \to B^\dagger + \lambda$, while
keeping $B$ fixed. The solution is thus
\be\label{Xab}
X(\alpha,\beta) = \l(e^\alpha + e^\beta - e^{\alpha+\beta}\r)^{-\lambda}
e^{ -\log\l(1-e^\alpha+ e^{\alpha-\beta}\r)
B^\dagger} \ket{0}.
\ee
Having found this general formula we shall now specialize to the
wedge states in the combined matter and ghost CFT with vanishing
central charge and vanishing anomaly $\lambda=0$. The separate
matter or ghost parts will be treated in subsection
\ref{MatterWedges}. Applying formula (\ref{Xab}) for the finite
conformal transformation operators (\ref{Udef}) we get
\be\label{UUdagger}
U_r U_s^\dagger \ket{0} = U_{2+\frac{2}{r}(s-2)}^\dagger \ket{0}.
\ee
As a check, let us note that for the special cases $r=2$ or $s=2$
it gives the correct result, less trivially it is also compatible
with the composition rule (\ref{comprule}).

Composition (\ref{UUdagger}) can be easily obtained using the gluing
theorem.\footnote{I would like to thank A.~Sen for this suggestion.}
Take an arbitrary state $\bra{\phi}$ and calculate both
\be
\bra{\phi} U_r U_s^\dagger \ket{0}
\ee
and
\be
\bra{\phi} U_t^\dagger \ket{0},
\ee
The latter is a path integral over a Riemann surface made by
gluing a half-disk with $\phi$ insertion and a piece of helix of
total opening angle $\pi (t-1)$. It is a cone of opening angle
$\pi t$. The former inner product instead, is a path integral over
the glued surface of two helices with angles $\pi (r-1) $ and
$\pi (s-1)$. The first helix has an insertion of transformed
$\phi$. After the gluing we transform back by a map $z \to
z^{2/r}$ to have a normal insertion of $\phi$. So we end up with a
integral over a cone with opening angle $\frac{2\pi}{r} ( r+s-2)$.
Matching
\be
\frac{2\pi}{r} ( r+s-2) = \pi t
\ee
we obtain
\be
t= 2+\frac{2}{r}(s-2),
\ee
which gives precisely the relation we found above quite
laboriously using the Virasoro algebra.

Taking the derivative of (\ref{UUdagger}) with respect to $r$ at $r=2$, or just
directly from (\ref{BeBdagger}) we find
\be
A U_s^\dagger \ket{0} = \frac{2-s}{s} A^\dagger U_s^\dagger
\ket{0}.
\ee
We may now introduce an important operator
\be
D \equiv A - A^\dagger = \sum_{k=1}^\infty
\frac{(-1)^k}{(2k-1)(2k+1)} (L_{2k} - L_{-2k}),
\ee
which acts as an exterior derivative of the star product algebra
because it is a linear combination of the operators $K_{2k}=
L_{2k} - L_{-2k}$ \cite{GJ2,RZ}. Its action on a wedge state is
\be
D U_s^\dagger \ket{0} =  \l(\frac{2}{s} -2 \r) A^\dagger
U_s^\dagger \ket{0}.
\ee
It obviously vanishes for the identity wedge state with $s=1$, but
it also annihilates the sliver with $s=\infty$, since
\be
A^\dagger  U_\infty^\dagger \ket{0} = 0
\ee
is just the $L_0$ conservation law for the sliver \cite{RSZ2}.

By the same method as we used to get (\ref{UUdagger}) we can also calculate
\be\label{eUwedge}
e^{-\alpha D}  U_s^\dagger \ket{0} = U_{1+e^\alpha (s-1)}^\dagger
\ket{0}.
\ee
It says, that starting with any regular wedge state $\ket{s}$,
where $1<s<\infty$ we can obtain any other wedge state by a finite
reparametrization $e^{-\alpha D}$. In the limit $\alpha \to
-\infty$ we then recover the identity wedge state $\ket{I}$ and in
the other limit  $\alpha \to +\infty$ we get the sliver. The
conclusion is that the identity and the sliver are just singular
reparametrizations of the ordinary vacuum state. This is in
complete agreement with the geometric arguments of \cite{GRSZ}.

To end up this discussion we would like to note that
(\ref{UUdagger}) and (\ref{eUwedge}) can be written in more
generality as operator statements
\bea\label{UUdagOP}
U_r U_s^\dagger &=& U_{2+\frac{2}{r}(s-2)}^\dagger
U_{2+\frac{2}{s}(r-2)},
\nonumber\\
e^{-\alpha D} &=& U_{1+e^\alpha}^\dagger U_{1+e^{-\alpha}}.
\eea
It follows by considering the BPZ or hermitian conjugation and from
the fact that the right hand side should be expressible entirely in
terms of $A$ and $A^\dagger$.

\subsection{Star products of wedge states without gluing theorem}

From the definition of the star product one can easily obtain
formulas for star product of vacuum state with any other state
from the Fock space
\bea
\ket{0}*\ket{\psi} &=& U_3^\dagger e^{-\frac{1}{\sqrt{3}} L_{-1}}
\l(\frac{4}{3}\r)^{L_0} e^{-\frac{1}{\sqrt{3}} L_{1}} U_3
\ket{\psi},
\nonumber\\
\ket{\psi}*\ket{0} &=& U_3^\dagger e^{\frac{1}{\sqrt{3}} L_{-1}}
\l(\frac{4}{3}\r)^{L_0} e^{\frac{1}{\sqrt{3}} L_{1}} U_3
\ket{\psi}.
\eea
These formulas make perfectly sense in the level expansion, since if
$\ket{\psi}$ contains finitely many levels, the whole expression
can be calculated to any given level exactly, in finitely many steps.
We have also checked it independently on several examples.

For the applications to the wedge states it is however convenient
to rewrite them using the formulas
\bea
e^{\alpha K_1} &=& e^{\tan\alpha\, L_{-1}}\,
 \l(\cos\alpha\r)^{-2 L_0 } \, e^{\tan\alpha\, L_{1}},
\nonumber\\
e^{\alpha K_1} e^{\beta A} &=& e^{\beta A} e^{\alpha\, e^{\frac{\beta}{2}} K_1},
\eea
to get
\bea\label{star0psi}
\ket{0}*\ket{\psi} &=& U_3^\dagger e^{-\frac{\pi}{6} K_1} U_3
\ket{\psi} = U_3^\dagger U_3 e^{-\frac{\pi}{4} K_1} \ket{\psi},
\nonumber\\
\ket{\psi}*\ket{0} &=& U_3^\dagger e^{+\frac{\pi}{6} K_1} U_3
\ket{\psi}= U_3^\dagger U_3  e^{+\frac{\pi}{4} K_1} \ket{\psi}.
\eea
From (\ref{K1Acomm}), (\ref{K1conserv}) and (\ref{UUdagger}) it follows that
\footnote{Note that in our notation $\ket{0} \equiv \ket{2}$.}
\bea
\ket{0}*\ket{r} &=& \ket{r+1},
\nonumber\\
\ket{r}*\ket{0} &=& \ket{r+1}.
\eea
Star product of two general wedge states can be calculated using
the formula (\ref{eUwedge}) and using the property
\be\label{transfstar}
e^{-\alpha D} \l( \ket{\phi} * \ket{\chi} \r) = e^{-\alpha D}
\ket{\phi} *  e^{-\alpha D} \ket{\chi} \qquad \forall\, \phi,
\chi,
\ee
valid for any derivative $D$ of the star product. Writing the
wedge states as
\bea
\ket{r} &=& e^{-\log(r-1)\, D} \ket{0},
\nonumber\\
\ket{s} &=& e^{-\log(r-1)\, D} \ket{1+\frac{s-1}{r-1}},
\eea
we easily find
\bea
\ket{r}*\ket{s} &=& e^{-\log(r-1)\, D} \ket{2+\frac{s-1}{r-1}} =
\nonumber\\
 &=& \ket{r+s-1}.
\eea

\subsection{On the matter and ghost parts of wedge states }
\label{MatterWedges}

Since all the wedge states are exponentials of total Virasoro operators,
they are naturally factorized into a matter and ghost parts. What
are the properties of these parts? What happens when the matter part
corresponds to a different wedge state than the ghost part? 
We will see that our techniques can be used to get some insight into
these questions.

By formal replacement $A \to A - \frac{\lambda}{4}$ and 
$A^\dagger \to A^\dagger - \frac{\lambda}{4}$ in 
(\ref{UUdagOP}) we can derive important relations 
valid in a CFT with  nonzero central charge $c$
\bea\label{UUdacom}
U_r U_s^\dagger &=& \l(\frac{rs}{2(r+s-2)}\r)^{\lambda}
U_{2+\frac{2}{r} (s-2)}^\dagger U_{2+\frac{2}{s} (r-2)},
\nonumber\\
e^{-\alpha D} &=& \l(\cosh\frac{\alpha}{2}\r)^{-\lambda}
U_{1+e^\alpha}^\dagger U_{1+e^{-\alpha}},
\eea
and the analog of (\ref{eUwedge}) is
\be\label{eDwedgematt}
e^{-\alpha D}  U_s^\dagger \ket{0} =
\l[\frac{1+e^\alpha (s-1)}{s e^{\frac{\alpha}{2}}}\r]^{-\lambda}
U_{1+e^\alpha (s-1)}^\dagger \ket{0}.
\ee
How can we now calculate the star product? Let us start with the
simplest one $\ket{0}*\ket{r}$. Using the explicit formula
(\ref{star0psi}) we get 
\be\label{0rmatt}
\ket{0}*\ket{r} = \l(\frac{3r}{2(r+1)}\r)^{\lambda} \ket{r+1}.
\ee
Note that this result is consistent with cyclicity 
\be\label{cycl}
\aver{A, B*C} = \aver{B, C*A} = \aver{C, A*B},
\ee
since applying $\bra{0}$ to the left hand side of (\ref{0rmatt}) we
find using (\ref{UUdacom})
\be
\bra{0} (\ket{0}*\ket{r}) = \aver{r|3} =
\l(\frac{3r}{2(r+1)}\r)^{\lambda},
\ee
the same as by applying it to the right hand side.

To calculate star product of arbitrary two wedge states we start from 
\be
\ket{r} * \ket{s} = e^{\aa_w} \ket{r+s-1},
\ee 
where $\aa_w$ is some anomaly in this composition rule. That the
product should take this form in general we know from the
factorization of the three vertex and from the composition of wedges
in combined matter and ghost CFT with $c=0$. The anomaly can again be
determined by contracting the equation with the vacuum $\bra{0}$, 
using the cyclicity  (\ref{cycl}) and equations (\ref{0rmatt}) 
(\ref{UUdacom}). The result is 
\be
e^{\aa_w} =  \l(\frac{3rs}{4(r+s-1)}\r)^{\lambda}.
\ee
Combining now the wedge states in the matter and ghost CFT 
into $\ket{r,\tilde r} = \ket{r}_m \otimes \ket{\tilde r}_{gh}$
we find
\be\label{unbalwedgecomp}
\ket{r,\tilde r} * \ket{s,\tilde s} =
\l(\frac{3rs}{4(r+s-1)}\r)^{\lambda}
\l(\frac{3\tilde r \tilde s}{4(\tilde r+\tilde
s-1)}\r)^{\tilde\lambda} 
\ket{r+s-1,\tilde r + \tilde s -1}.
\ee 
Note, that we allow for arbitrary central charges and therefore
anomalies $\lambda, \tilde \lambda$ in both CFT's. This is important
for applications to the twisted ghost CFT in \cite{GRSZ}.

The anomalous factors in (\ref{unbalwedgecomp})  may look at first 
rather odd, but in fact they are nicely compatible with commutativity
and associativity. 

Now let us study what happens under finite or infinitesimal
reparametrizations  generated by 
\be
D^X = \sum_{k=1}^\infty
\frac{(-1)^k}{(2k-1)(2k+1)} K_{2k}^{X},
\ee
where the superscript $X$ refers to matter part of the CFT.
On general grounds of cyclicity one can argue that the anomaly
$\aa$ appearing in  
\be\label{transfstarmattinf}
D^X \ket{\phi} *  \ket{\chi} + \ket{\phi} *  D^X \ket{\chi}=
(\aa +D^X ) \l(\ket{\phi} * \ket{\chi} \r), 
\ee
is the same for all the unbalanced wedge
states. The integrated form is then 
\be\label{transfstarmatt}
e^{-\alpha D^X} \ket{\phi} *  e^{-\alpha D^X} \ket{\chi} =
e^{-\alpha \aa} e^{-\alpha D^X} \l( \ket{\phi} * \ket{\chi} \r). 
\ee
Since we know explicitly from (\ref{eDwedgematt}) that 
\be
e^{-\alpha D^X} \ket{r,\tilde r} = \l[\frac{1+e^\alpha (r-1)}{r e^{\frac{\alpha}{2}}}\r]^{-\lambda}
\ket{1+e^\alpha (r-1), \tilde r}
\ee 
we may calculate both sides of (\ref{transfstarmatt}) to find the
anomaly
\be
\aa = - \frac{\lambda}{2}.
\ee
An alternative procedure is to calculate the anomaly directly
by summing the anomalies associated to individual $K_{2k}^{X}$
calculated in \cite{GJ2,Romans,RZ}
\be
\aa=\sum_{k=1}^\infty \frac{(-1)^k}{(2k-1)(2k+1)} \cdot (-3)
\frac{5c}{54} \cdot k (-1)^k = -\frac{5}{12} \lambda.
\ee
Although the numbers $-\frac{1}{2}$ and $-\frac{5}{12}$ are quite
close, we got a clear clash between these two equally justifiable
methods.\footnote{Further clash arises when we want to consistently
include the identity $\ket{1,1}$. This would require 
$\aa =-\frac{3}{4} \lambda$.} 
We have to conclude that our basic assumption of the
cyclicity is wrong. If we include the unbalanced wedge states into the
\sf algebra and/or we allow for reparametrization generated by $D^X$
we definitely violate the cyclicity of the three vertex.  

Let us end this section with another remark. From relation
(\ref{UUdacom}) follows that the norm squared of the matter part of wedge
states is
\be
\aver{r|r} = \l(\frac{r^2}{4(r-1)}\r)^{\lambda}.
\ee
It is divergent for all $r$. In particular the matter part of the
state $\ket{3}$, which is a star product of two vacuum states has
infinite norm. This was recently found also in \cite{Moore}.
The norm of
the identity $r=1$ and of the sliver $r=\infty$ is even more
divergent, it diverges even at finite $\lambda$.

\section{Identity string field}
\label{Identity}

In this section we would like to turn our attention to the identity element of the
\sf algebra. In general, identity element of any algebra is quite an important object.
It may or may not exist. For the \sf star algebra we shall give an explicit construction
bellow. However, since we are lacking a mathematically satisfactory definition of the
algebra itself we cannot say whether the identity actually belongs to the space or not.
Good definition of the algebra would be to require having only finite
norm states, for instance, but this only shifts the problem to finding
a good norm. The canonical norm defined by the hermitian inner
product does not work, since as we have seen just above in section
\ref{MatterWedges} even the product of two vacuum
states $\ket{0}$ does not have finite norm.

Let us now forget about the problems whether the identity should
belong to the algebra or not and let us describe its various
forms. In the Witten's formulation where the star product of two
string fields in the Schr\"odinger representation is defined by
(ignoring ghosts for the moment)
\bea\label{starprodW}
&& (\Psi_1 * \Psi_2)(X_0(\sigma))=
 \int \dd X_1(\sigma) \dd X_2(\sigma) \, \Psi_1(X_1(\sigma))\Psi_2(X_2(\sigma))
\\\nonumber
&& \quad \prod_{0\leq\sigma\leq\frac{\pi}{2}}
\delta(X_1(\sigma)-X_0(\sigma))
\delta(X_2(\sigma)-X_1(\pi-\sigma))
\delta(X_0(\pi-\sigma)-X_2(\pi-\sigma)),
\eea
the identity is clearly the functional
\be
\aver{X(\sigma)|I} = \prod_{0\leq\sigma\leq\frac{\pi}{2}}
\delta(X(\sigma)-X(\pi-\sigma)).
\ee
To write it in the Fock space, first one need to use the mode expansion
$X(\sigma)=x_0+\sqrt{2}\,\sum_{n=1}^{\infty} x_n \cos n\sigma$
to get
\be
\aver{X(\sigma)|I} = \prod_{n=1,3,\ldots} \delta(x_n).
\ee
Then after expressing the coherent states $\ket{x_n}$ using the creation operators
we find by a simple calculation
\bea
\ket{I} &=& \int \dd X(\sigma)\, \ket{X(\sigma)} \aver{X(\sigma)|I}
\nonumber\\
&=& e^{-\frac12 \sum_{n=0}^\infty (-1)^n a_n^\dagger a_n^\dagger} \ket{0}.
\eea
Treating carefully the ghosts one gets in the oscillator approach
\cite{GJ2}
\be\label{identityGJ}
\ket{I} = \frac i4 b(i) b(-i)
e^{ \sum_{n=1}^\infty (-1)^n
(-\frac12 a_n^\dagger a_n^\dagger + c_{-n} b_{-n})} c_1 c_0 \ket{0}.
\ee
From the geometrical representation of the star product
discussed in section \ref{wedgestates} it is clear that the identity
is the wedge state $\ket{1}$
\be\label{identityRZ}
\ket{I} = e^{L_{-2} - \frac12 L_{-4} + \frac12 L_{-6} - \frac{7}{12} L_{-8}
+ \frac{2}{3} L_{-10} - \frac{13}{20} L_{-12}+ \cdots}\ket{0}.
\ee
We have calculated the higher level terms in the exponent exactly up to
$L_{-100}$ term, the results are plotted in the graph Fig. \ref{FigId}.
It is quite surprising that up to the level 20 the coefficients are less or
around one, but then start to diverge faster than any exponential.
This divergence should be however viewed as some combinatorial divergence
related to unfortunate ordering of the Virasoro generators. Indeed, a nice
alternative form of the identity has been found by Ellwood et. al.
\cite{EFHM}
\be\label{identityEFHM}
\ket{I} = \l(\prod_{n=2}^\infty e^{-\frac{2}{2^n} L_{-2^n}}\r) e^{L_{-2}}\ket{0},
\ee
in which higher level terms have manifestly well behaved coefficients.

Finally let us note, that one can easily perform an explicit calculation in level truncation
to show, that various forms of the identity (\ref{identityGJ}), (\ref{identityRZ})
and (\ref{identityEFHM}) are indeed in mutual agreement.

\subsection{Conservation laws for the identity}

\subsubsection*{Virasoro conservation laws}

Let us recall the derivation of the conservation laws
due to Rastelli and Zwiebach \cite{RZ}.
We start with a global coordinate $\tilde z$ on the 1-punctured disk, associated
to the identity $\bra{I}$. For any holomorphic vector field $\tilde v(\tilde z)$
we have the basic identity
\be\label{Tconsbasic}
\bra{I} \oint_{\ccc} d\tilde z\, \tilde v(\tilde z) \tilde T(\tilde z) = 0,
\ee
where $\ccc$ is a contour encircling the puncture.
Passing to the local coordinate $z$ around the puncture we get
\be
\bra{I} \oint_{\ccc} dz\, v(z) \l( T(z) - \frac{c}{12} S(f^{360^\circ}(z),z) \r) = 0,
\ee
where
\be
S(f^{360^\circ}(z),z) = 6 (1+z^2)^{-2} = 6 \sum_{m=1}^\infty m(-1)^{m-1}
z^{2(m-1)}
\ee
is the Schwarzian derivative reflecting the non-tensor character of the energy
momentum tensor when the central charge $c$ is nonzero.
For a particular choice of the vector field
$v(z)=z^{n+1} - (-1)^n z^{-n+1}$, which is holomorphic everywhere
in the global coordinate $\tilde z$ except the puncture, we get
\bea
K_{2n} \ket{I} &=& -\frac{c}{2}\, n (-1)^{n} \ket{I},
\nonumber\\
K_{2n+1} \ket{I} &=& 0,
\eea
where we define
\be
K_n = L_n - (-1)^n L_{-n}.
\ee
The same identities can be derived for the $b$ ghost, in that case, there is no
anomaly however.

Let us further comment on some applications of the formulas we have obtained.
First one can rewrite the state $T(z) \ket{I}$ in a form which is manifestly
well defined in the level expansion and perform the geometric sums provided
$|z| >1$.
\be\label{TzI}
T(z) \ket{I} = \frac c2 \frac{1}{(1+z^2)^2} \ket{I} +
\frac{1}{z^2} L_0 \ket{I} + \frac{1}{z^2} \sum_{n \ge 1}
\l(z^n + (-1)^n z^{-n}\r) L_{-n} \ket{I}.
\ee
From these identities, and those for the $b$ ghost, one can easily check the
overlap equations
\bea
\l(T(z)-\frac{1}{z^4} T(-1/z)\r) \ket{I} &=& 0,
\nonumber\\
\l(b(z)-\frac{1}{z^4} b(-1/z)\r) \ket{I} &=& 0.
\eea

\subsubsection*{Conservation of the $c$-ghost}

We start from the identity
\be\label{cconsbasic}
\bra{I} \oint_{\ccc} dz\, \phi(z) c(z) = 0,
\ee
where $\phi(z)$ is a quadratic differential holomorphic everywhere except at the
puncture located at the origin, and $\ccc$ is a contour encircling the puncture.
The $\phi(z)$ transforms as follows
\be
\tilde\phi(\tilde z) = \l(\frac{dz}{d\tilde z}\r)^2 \phi(z).
\ee
We shall pass from the local coordinate $z$ around the puncture to the
global coordinate on the 1-punctured disk
\be
\tilde z = \frac{2z}{1-z^2}.
\ee
For the particular choice of the quadratic differentials
\bea
\phi_{2n}(z) &=& \frac{1}{z^2} \l(z^n - \l(-\frac1z \r)^n \r)^2,
\nonumber\\
\phi_{2n+1}(z) &=& \frac{1}{z^2}
\l(z^{2n+1} - \l(-\frac1z \r)^{2n+1} - (-1)^n \l(z-\frac1z\r) \r),
\eea
where $n \ge 1$, the transformed differentials are
\bea
\tilde\phi_{2n}(\tilde z) &=& \frac{4}{{\tilde z}^{2n+2}}
\l(\sum_{k=1,3,5,\ldots}
\l(\begin{array}{c} n \\ k \end{array}\r) \l(1+{\tilde z}^2 \r)^{\frac{k-1}{2}} \r)^2,
\\\nonumber
\tilde\phi_{2n+1}(\tilde z) &=& -\frac{2}{{\tilde z}^{2n+3}} \frac{1}{1+{\tilde z}^2}
\l(-(-1)^n {\tilde z}^{2n} +\sum_{k=0,2,4,\ldots}
\l(\begin{array}{c} 2n+1 \\ k \end{array}\r) \l(1+{\tilde z}^2
\r)^{\frac{k}{2}} \r).
\eea
All the sums here are finite due to the combinatorial factors which are defined to
be zero whenever the lower entry is bigger than the upper entry.

The quadratic differentials expressed in the global coordinate system $\tilde z$
are holomorphic in the whole complex plane except zero, in particular they do
not have any singularity at $\pm i$. Therefore one may derive
from (\ref{cconsbasic}) the conservation laws\footnote{For a
derivation using oscillators see \cite{TT}.}
\bea
C_{2n} \ket{I} &=& (-1)^n C_0 \ket{I},
\nonumber\\
C_{2n+1} \ket{I} &=& (-1)^n C_1 \ket{I}.
\eea
where in general we define
\be
C_k = c_k + (-1)^k c_{-k}.
\ee
Let us remark that a naive conservation law based on the quadratic differential
\be
\phi(z) = \frac{1}{z^2} \l(z^n - \l(-\frac1z \r)^n \r)
\ee
fails, since $\tilde\phi(\tilde z)$ does have poles in $\pm i$. This is
actually a simple manifestation of the midpoint anomalies.

Again, as we have done for the energy momentum tensor,
we can rewrite the state $c(z) \ket{I}$ in a form which is manifestly
well defined in the level expansion provided $|z| >1$,
\be\label{caI}
c(z) \ket{I} = -z \frac{1-z^2}{1+z^2}\,c_0 \ket{I} +
\frac{z^2}{1+z^2}\l(c_1-c_{-1}\r) \ket{I} + z \sum_{n \ge 1} \l(z^n -
\l(-\frac1z\r)^n\r)c_{-n} \ket{I}.
\ee
The single poles for $z \to \pm i$ were first found by other means by
\cite{QS,HM}.  From this formula, it is a simple exercise to verify the
overlap equation
\be
\l(c(z)-z^2 c(-1/z)\r) \ket{I} = 0.
\ee
Another observation we can make is about $c_L \ket{I}$, where
\be
c_L = \frac12 c_0 + \frac{1}{\pi} \sum_{k=0}^\infty \frac{(-1)^k}{2k+1}
\l(c_{2k+1}+c_{-(2k+1)}\r).
\ee
We can easily calculate that
\be
c_L \ket{I} = \frac12 c_0 \ket{I} +  \frac{2}{\pi} \sum_{k=0}^\infty \frac{(-1)^k}{2k+1}
c_{-(2k+1)} \ket{I}  +  \frac{1}{\pi} \sum_{k=0}^\infty \frac{1}{2k+1} C_1 \ket{I},
\ee
which is divergent due to the last term. This fact rules out the possibility
of relating solutions to the vacuum and Witten's cubic \sf theories through
the most naive $(c_L -Q_L) \ket{I}$ shift. For more sophisticated
possibilities see \cite{KO}.

\subsubsection*{Current conservation laws}

For completeness let us also consider conservation laws for currents.
Let us take in general a holomorphic current $J(z)$ having the following
OPE with the stress tensor
\be
T(z) J(0) = \frac{2q}{z^3} + \frac{J(0)}{z^2} +
\frac{\partial J(0)}{z}.
\ee
Under finite conformal transformations it transforms as
\be
\frac{dw}{dz} \tilde J(w) = J(z) - q \frac{d^2 w}{dz^2} \l(\frac{dw}{dz}\r)^{-1}.
\ee
The anomalous constant $q$ is zero for the BRST current and $\partial X$
currents, for the ghost number current it is $-\frac32$.
Following the same procedure as above, one may derive
conservation laws
\bea
H_0 \ket{I} &=& 0,
\nonumber\\
H_{2k+1} \ket{I} &=& 0,
\nonumber\\
H_{2k} \ket{I} &=& (-1)^k 2q \ket{I},
\eea
where
\be
H_k = J_k + (-1)^k J_{-k}.
\ee

\subsection{Anomalous properties of the identity}

One particularly puzzling aspect \cite{RZ,KO} of the identity is the following:
We know that $c_0$ acts as a derivation on the \sf algebra. Therefore we can
write formally
\bdm
c_0 \ket{\Psi}
= c_0 \l( \ket{I} * \ket{\Psi} \r)
= c_0 \ket{I} * \ket{\Psi}  +  \ket{I} * c_0\ket{\Psi}
= c_0 \ket{I} * \ket{\Psi}  +  c_0\ket{\Psi},
\edm
from which follows that
\be
c_0 \ket{I} * \ket{\Psi} = 0
\ee
for any \sf $\ket{\Psi}$. Naively one would conclude (taking
$\ket{\Psi}=\ket{I}$)  that $c_0 \ket{I} = 0$, but
that is manifestly not true. One could possibly imagine several ways out:
\begin{itemize}
\item
$\ket{I}$ is not a true identity on all states,
\item
$c_0$ is not a true derivation on the whole algebra
\item
even though  $c_0 \ket{I} \ne 0$, still we have $ c_0 \ket{I} * \ket{\Psi} = 0$ for any
'well behaved' state $\ket{\Psi}$
\item
simply $ c_0 \ket{I} * \ket{\Psi}$ is not well defined in the level expansion
\end{itemize}
We will argue for the last possibility, but in some limited sense all the
explanations are true.

The derivations of the fact that $\ket{I}$ is the identity and that $c_0$ is a
derivation on the algebra are quite firm when one restricts on well behaved
states. Again it is difficult to say what is a well defined state, but those
which contain finitely many levels certainly are. We have checked numerically
that the identity is an identity on many states, it seems that it is an identity
even on itself and other wedge states, which might be otherwise somehow problematic.

To check the third possibility it is best to look first at some example.
Let us calculate
\bdm
c_0 \ket{I} * \ket{0}.
\edm
In general there are many ways to do the calculation. The most naive way would
be to truncate the identity to some maximal level, legally use the $c_0$
conservation to arrive at  $c_0 (\ket{I} * \ket{0}) -  \ket{I} * c_0 \ket{0}$
which is indeed very close to zero, since the identity $\ket{I}$ works well for
the states $\ket{0}, c_0 \ket{0}$. A sort of 'canonical' way of calculation
suggested in \cite{RZ} is to re-order $c_0 \ket{I}$ to have only the ghost $c_1$
acting on the vacuum and Virasoro generators acting on it from the left. Then
one can use the recursive relations of \cite{GJ2,Samuel,RZ} for the Virasoro
generators, to reduce the expression to the linear combination of terms
\bdm
L_{-a} L_{-b} \ldots (c_1 \ket{0} * \ket{0}).
\edm
Actually we can perform this calculation exactly even with some sort of
regularization. As a first step let us commute the $c$-ghost to the vacuum.
In more generality, we will do it for an arbitrary wedge state instead
of the identity and for convenience work with the bra vectors
\be
\bra{r} c_0 = \bra{0} U_r c_0 =  \bra{0} \l(U_r c_0 U_r^{-1}\r) U_r.
\ee
The factor in the bracket is readily
\bea\label{c0transf}
U_r c_0 U_r^{-1} &=& \oint \frac{dz}{2\pi i} \frac{1}{z^2}\, U_r c(z) U_r^{-1}
\nonumber\\
 &=& -4 r^2 \sum_n c_n \oint \frac{dw}{2\pi i} \frac{1}{w^{n-1}}
\frac{(1+w^2)^{r-2}}{\l[(1+iw)^r - (1-iw)^r\r]^2}
\\\nonumber
 &=& c_0 + \frac{r^2-4}{3} c_2 + \frac{r^4-10 r^2 + 24}{15} c_4 +
\frac{10 r^6 -168 r^4 + 945 r^2 -1732}{945} c_6 + \cdots.
\eea
For general $r$ it looks difficult to find a closed expression, for the identity
state which corresponds to $r=1$ the sum can be easily performed
\be
U_1 c_0 U_1^{-1} = -\frac{i}{2} \l(c(i) - c(-i)\r).
\ee
Therefore
\be
c_0 \ket{I} = \frac{i}{2}\, U_1^\dagger \l(c(i) - c(-i)\r) \ket{0}.
\ee
This is already in the form for which we know how to calculate the star products
exactly. Recall however that the insertion points $\pm i$ are singular. A
natural regularization from the point of view of level expansion is to replace
\be
c(i) - c(-i) \quad \longrightarrow \quad c(ai) - c(-ai),
\ee
where $a<1$ is approaching unity from bellow.\footnote{Another interesting regularization
would be to keep $r \ne 1$, but that is technically rather cumbersome, due to the
presence of nontrivial contour integral in (\ref{c0transf}). Yet another
possibility is to replace $U_1$ with $U_r$ for $r \ne 1$ in (\ref{c0Isgen}),
this is however an almost trivial modification of our calculations.}
Let us calculate in more generality
\bea\label{c0Isgen}
\frac{i}{2}\, U_1^\dagger \l(c(ia) - c(-ia)\r) \ket{0} *  U_s^\dagger  \ket{0} &=&
\frac{is}{2} \frac{1-a^2}{1+{x'}^2(ia)} U_s^\dagger c(x'(ia)) \ket{0}
\\\nonumber\
&&
-\frac{is}{2} \frac{1-a^2}{1+{x'}^2(-ia)} U_s^\dagger c(x'(-ia)) \ket{0},
\eea
where
\be
x'(ia) = -i \frac{(a-1)^{\frac2s} + (a+1)^{\frac2s}}{(a-1)^{\frac2s} - (a+1)^{\frac2s}}.
\ee
This has well defined limit for $a \to \pm 1$, it is either $i$ or $-i$
respectively. The prefactors require little bit more care, since they are
limits of $0/0$ type.

For $s>2$, the whole expression is well defined and we get
\be
\lim_{a\to 1^{-}} \frac{i}{2}\, U_1^\dagger \l(c(ia) - c(-ia)\r) \ket{0} *
U_s^\dagger  \ket{0} = 0.
\ee
For $s=2$ we get
\be
\lim_{a\to  1^{-}} \frac{i}{2}\, U_1^\dagger \l(c(ia) - c(-ia)\r) \ket{0} *
\ket{0} = i (c(i) - c(-i)) \ket{0}.
\ee
For $s<2$ at least one of the two prefactors in (\ref{c0Isgen}) is divergent.
The conclusion of the above calculation is that the result of the calculation
$c_0 \ket{I} * \ket{0}$ is highly sensitive to the method used. From the
mathematical point of view, $c_0 \ket{I}$ does not belong to the star algebra.
If we want to have the derivation $c_0$ defined on all of the algebra, we should
conclude, that neither the identity $\ket{I}$ belongs to the
algebra. Alternatively we can think of the identity belonging to the algebra,
but then the derivative $c_0$ maps some elements of the algebra out of it.

One could imagine other ways how to study the object $c_0 \ket{I}$. We would
like to warn the reader of the following problem
\bea
\l[\oint \frac{dz}{2\pi i} \frac{c(z)}{z^2}\r] \ket{I} &=& c_0 \ket{I},
\nonumber\\
\oint \frac{dz}{2\pi i} \l[ \frac{c(z)}{z^2} \ket{I} \r] &=& - c_0 \ket{I},
\eea
where the brackets in the second case mean, that we are evaluating the
expression, i.e. calculating it in the level expansion through formula (\ref{caI}).
Both integrals are along small contours around the origin.
To get identical results, we would need to include the points $\pm i$ inside the
contour of integration. The problem can be traced back to the fact, that to make
sense of $c(z)\ket{I}$ without analytic continuation, we need to remain outside
of the unit circle.

Finally we would like to address the issue of the 'integrated' anomaly.
Let us consider
\be\label{consVcI}
\bra{V_{123}} (c_0^{(1)}+c_0^{(2)}+c_0^{(3)}) \ket{I}\otimes\ket{\Psi_2}\otimes\ket{\Psi_3},
\ee
where for simplicity $\ket{\Psi_{2,3}}$ are two ghost number one string fields.
This is equal to
\bea
&& \bra{\bpz \Psi_3}\, \biggl(c_0\ket{I} * \ket{\Psi_2} + \ket{I} * c_0 \ket{\Psi_2} -c_0
\l(\ket{I} * \ket{\Psi_2}\r) \biggr)
\nonumber\\
&& = \bra{\bpz \Psi_3}\, \biggl( c_0\ket{I} * \ket{\Psi_2}\biggr),
\eea
which we call 'integrated' anomaly. In level truncation we can safely use the
cyclicity to get further
\be
-\bra{I} c_0 \, \biggl(\ket{\Psi_2}*\ket{\Psi_3}\biggr) = \aver{\bpz \Psi_3 |
c_0 | \Psi_2} - \aver{\bpz c_0 \Psi_3 | \Psi_2} = 0.
\ee
This is as it should be, since our starting expression
(\ref{consVcI}) is also manifestly zero in the level expansion.
The moral is that while $c_0\ket{I} * \ket{\Psi_2}$ itself is ill
defined, its BPZ inner products with well behaved states can be
consistently set to zero.

\subsection{Application to the tachyon condensation}

There is one remarkably simple application of the above results to the study of
tachyon condensation in ordinary cubic \sf theory.
The string field action at the critical point should
be invariant under a general variation
\be\label{varS}
\delta S = -\frac{1}{g^2} \biggl[ \aver{\delta\Phi | Q \Phi_0} + \aver{\delta \Phi|
\Phi_0 * \Phi_0} \biggr] = 0.
\ee
Taking a variation of the form $\delta \Phi = C_1 \ket{I}$ we see, that
the variation of the cubic term vanishes since $C_1$ like $c_0$ is a
derivation of the star algebra and because the identity commutes with everything.
Therefore we are led to the conclusion that the tachyon condensate should satisfy
\be\label{linid}
\aver{I|C_1 Q |\Phi_0} = 0.
\ee
Let us see how well this works in the level truncation.
We start our analysis with a level 10 numerical solution $\Phi_0$ of
the equations of motion in Siegel gauge which has been kindly provided
to us by Gaiotto, Rastelli, Sen and Zwiebach.\footnote{Thanks also to
P.~J.~de~Smet for some technical help in manipulating the solution while
working together on another project.}
Now with this solution (treating it as an exact solution), we
calculate level $n=0, 2, 4, 6, 8$ approximation to the expression
$\aver{I|C_1 Q |\Phi_0}$ by truncating both the identity $\bra{I}$ and
$C_1 Q\ket{\Phi_0}$ to that level. We cannot go to level 10 in this
test, since this would require knowledge of level 12 terms in the solution.
The results are listed in the following table
\begin{displaymath}
\begin{array}{|r||c|c|c|c|c|}
\hline
{\rm Level:} & 0 & 2 & 4 & 6 & 8 \\ \hline
\aver{I|C_1 Q |\Phi_0} & 0.3294 & -0.1684 & 0.1303 & -0.0422 & 0.0671 \\
\hline
\end{array}
\end{displaymath}
Since $\aver{I|C_1 Q |\Phi_0}$ in a level expansion is a sum of many
terms which start with contribution of level 0 equal to
$t=0.5463\ldots$, we regard the values in the table as a reasonably good
confirmation of our claims.

One can be puzzled about the last entry
$0.0671$ which seems to grow again. This value however depends on the
values of level 10 components in the solution $\Phi_0$ which are
likely to be affected significantly by the fact that $\Phi_0$ itself
is found from the action truncated to level 10 fields at most.

\section{Concluding remarks}

One of the least explored aspects of the \sf algebra is the issue of 
associativity. It has been known for a long time that some star
products involving  operators integrated over half of the string 
violate associativity \cite{Horowitz,RT}.
In this paper we have seen two new occasions where the problems
might arise. First is when one keeps multiplying $L_{-2}\ket{0}$ by the
vacuum $\ket{0}$ from the right. In few steps the level truncation
will stop converging and associativity of the star product is recovered only after 
analytic continuation. Second case, where we definitely break the
associativity is in the case of unbalanced wedge states, whose matter
and ghost parts do not match. We have explicitly shown that in this
case the three vertex loses its cyclicity, which implies the loss of
associativity as well.  Alternatively, one could imagine that 
some anomalous conservation laws receive some additional corrections
when applied to wedge states, but this possibility is again in conflict
with associativity. This problem is likely not just an academic
question, since these anomalous conservation laws give nontrivial
information about the classical solutions of the \sft 
\cite{Schnabl2,Schnabl4}. 
Another type of anomaly is the famous $c_0 \ket{I}$ problem \cite{RZ}.
We have given a partial solution, but it would be nice to understand it
more deeply and in a broader context.

\section*{Acknowledgments}
I would like to thank L.~Bonora, P.~J.~de~Smet, N.~Moeller, L.~Rastelli,
A.~Sen, W.~Taylor and B.~Zwiebach for useful discussions. 
Part of this work has been excerpted from my PhD thesis written at
SISSA, Trieste and defended in October 2001. This work has
been also supported in part by DOE contract
\#DE-FC02-94ER40818.

\appendix
\section{Star products in level expansion }
\label{AppStarProds}

In this appendix we would like to collect some numerical results showing,
how well the level expansion works for star products. We have performed
some explicit checks at level 20, where one of the states was particularly simple,
and some other checks at level 16 which confirmed the composition law
(\ref{wedgecomp}) obtained by the gluing ideas described in section
\ref{wedgestates}. We have written for that purpose a computer program
in {\it Mathematica} which is based on the Virasoro conservation laws
for the three vertex \cite{Samuel,GJ2,RZ}.

\subsection{Some level 20 calculations }

First let us present the results, how good is the identity state
$\ket{I}=\ket{1}$ acting on some basic states
\begin{eqnarray}
\ket{0} * \ket{I} &=&
\ket{0}+0.00008L_{-2}\ket{0}-0.00007L_{-3}\ket{0}-0.00068L_{-4}\ket{0}+
 \nonumber\\ &&
+0.00039L_{-2} L_{-2}\ket{0} +\cdots \nonumber\\
L_{-2}\ket{0} * \ket{I} &=&
0.9987L_{-2}\ket{0}-0.0001L_{-3}\ket{0}-0.0001L_{-4}\ket{0}+
 \nonumber\\ &&
+0.0007L_{-2} L_{-2}\ket{0} +\cdots\nonumber\\
L_{-2}L_{-2} \ket{0} * \ket{I} &=&
0.0054L_{-2}\ket{0}-0.0001L_{-3}\ket{0}+0.0002L_{-4}\ket{0}+
 \nonumber\\ &&
+0.9967L_{-2} L_{-2}\ket{0} +\cdots\nonumber\\
L_{-4}\ket{0} * \ket{I} &=&
0.0035L_{-2}\ket{0}-0.0005L_{-3}\ket{0}+0.9967L_{-4}\ket{0}+
 \nonumber\\ &&
+0.0002L_{-2} L_{-2}\ket{0}+\cdots
\end{eqnarray}

Now let us present the results for the products of the vacuum $\ket{0}$
and other wedge states to verify the composition rule
$\ket{r} * \ket{s} = \ket{r+s-1}$.
\bea
\ket{2} * \ket{3} &=&
\ket{0}-0.25006L_{-2}\ket{0}+0.00197L_{-3}\ket{0}+0.03132L_{-4}\ket{0}+
 \nonumber\\ &&
+0.03126L_{-2} L_{-2}\ket{0} +\cdots\nonumber\\
\ket{2} * \ket{\infty} &=&
\ket{0}-0.32085L_{-2}\ket{0}+0.00563L_{-3}\ket{0}+0.03294L_{-4}\ket{0}+
 \nonumber\\ &&
+0.05137L_{-2} L_{-2}\ket{0} +\cdots\nonumber\\
\ket{2} * \ket{1/2} &=&
\ket{0}-38723.7L_{-2}\ket{0}-22117.4L_{-3}\ket{0}-12233.8L_{-4}\ket{0}+
 \nonumber\\ &&
+34414.4L_{-2} L_{-2}\ket{0} +\cdots
\nonumber\\
\eea
The first two products should be compared with the wedge states
\bea
\ket{4} &=&
\ket{0}-0.25L_{-2}\ket{0}+0.03125L_{-4}\ket{0}+0.03125 L_{-2} L_{-2}\ket{0}+\cdots
\nonumber\\
\ket{\infty} &=&
\ket{0}-0.33333L_{-2}\ket{0}+0.03333L_{-4}\ket{0}+0.05556L_{-2} L_{-2}\ket{0}+\cdots
\eea
We see that the agreement is quite good (within $0.23 \%$) for the
state $\ket{90^\circ}$  but is considerably worse (within $7.5 \%$) for the
$\ket{\infty}$.
The last product indicates that the states $\ket{r}$ with $r<1$ do not have much
sense in the level expansion.

\subsection{Level 16 calculations}

\begin{eqnarray}
\ket{I} * \ket{I} &=&
\ket{0}+1.00386L_{-2}\ket{0}-0.50098L_{-4}\ket{0}+ 0.49723L_{-2}
L_{-2}\ket{0} +\cdots
\nonumber\\
\ket{\infty}*\ket{\infty} &=&
\ket{0}-0.36150L_{-2}\ket{0}+0.03338L_{-4}\ket{0}+ 0.06549L_{-2}
L_{-2}\ket{0} +\cdots
\nonumber\\
\ket{\infty} * \ket{I} &=&
\ket{0}-0.32656L_{-2}\ket{0}+0.00267L_{-3}\ket{0}+0.03148L_{-4}\ket{0}+
 \nonumber\\ &&
+0.05365L_{-2} L_{-2}\ket{0} +\cdots
\nonumber\\
\ket{\infty} * \ket{3} &=&
\ket{0}-0.33434L_{-2}\ket{0}-0.00564L_{-3}\ket{0}+0.03394L_{-4}\ket{0}+
 \nonumber\\ &&
+0.05587L_{-2} L_{-2}\ket{0} +\cdots
\nonumber\\
\ket{1/2} * \ket{3} &=&
\ket{0}-2147.14L_{-2}\ket{0}+1327.72L_{-3}\ket{0}-553.046L_{-4}\ket{0}+
 \nonumber\\ &&
+2074.33L_{-2} L_{-2}\ket{0} +\cdots
\nonumber\\
\ket{3} * \ket{3} &=&
\ket{0}-0.28708L_{-2}\ket{0}+0.03348L_{-4}\ket{0}+ 0.04122L_{-2}
L_{-2}\ket{0} +\cdots
\nonumber\\
\ket{2} * \ket{3} &=&
\ket{0}-0.25008L_{-2}\ket{0}+0.00246L_{-3}\ket{0}+0.03135L_{-4}\ket{0}+
 \nonumber\\ &&
+0.03127L_{-2} L_{-2}\ket{0} +\cdots
\nonumber\\
\ket{2} * \ket{I} &=&
\ket{0}+0.00010L_{-2}\ket{0}-0.00008L_{-3}\ket{0}-0.00109L_{-4}\ket{0}+
 \nonumber\\ &&
+0.00066L_{-2} L_{-2}\ket{0} +\cdots
\nonumber\\
\ket{2} * \ket{\infty} &=&
\ket{0}-0.31966L_{-2}\ket{0}+0.00668L_{-3}\ket{0}+0.03293L_{-4}\ket{0}+
 \nonumber\\ &&
+0.05096L_{-2} L_{-2}\ket{0} +\cdots
\nonumber\\
\ket{2} * \ket{1/2} &=&
\ket{0}-1876.75L_{-2}\ket{0}-1163.92L_{-3}\ket{0}-567.608L_{-4}\ket{0}+
 \nonumber\\ &&
+1851.82L_{-2} L_{-2}\ket{0} +\cdots
\nonumber\\
\end{eqnarray}

Let us compare these results obtained in level expansion with the exact
answer. The errors for products with sliver state $\ket{\infty}$ are smallest
for $\ket{3}$ state: $0.3 \%$ at level 2 and $1.8 \%$ at level 4 coefficients.
The biggest error is when we multiply the sliver with another sliver 
$\ket{\infty}$. It is $8.4 \%$ at level 2 and $18 \%$ at level 4.

The errors for the product of $\ket{I}$ are again biggest for the
$\ket{\infty}$ state with $2.1 \%$ at level 2 and $5.6 \%$ at level 4.
The errors in the product of the identity with itself are $0.39 \%$ or
$0.55 \%$ respectively.

The errors for $\ket{0}*\ket{3}$ are the lowest of all of the
examples: $0.03 \%$ and $0.33 \%$ respectively.

The moral is that the wedge composition rule works better for states
closer to the vacuum. It works worse for the identity and the worst for
the $\ket{\infty}$ state.

\newpage

\section{Behavior of the wedge state coefficients}
\label{WedgeCoefficients}

\FIGURE{
\epsfbox{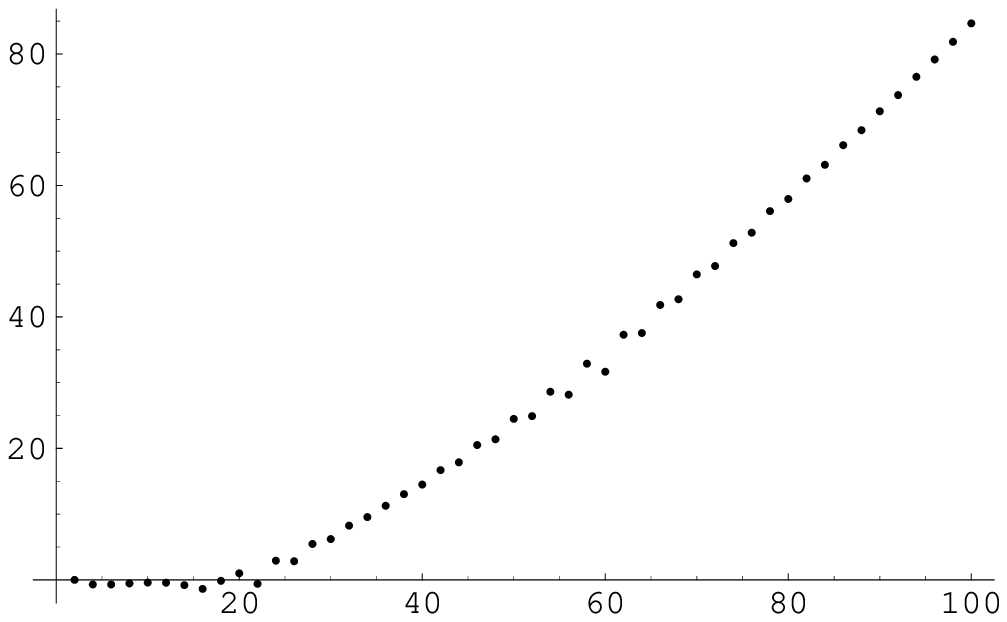}
\caption{\small Plot of $\log |v_{n}|$, where $v_n$ are coefficients 
appearing in the definition of the star algebra identity
$\ket{I}=\ket{1}$. The odd values of $n$ are omitted as all $v_{2k+1}=0$ 
trivially. Despite the apparent exponential growth of the
coefficients starting at level 20, the identity is well behaved in
level expansion.}
\label{FigId}
}

\FIGURE{
\epsfbox{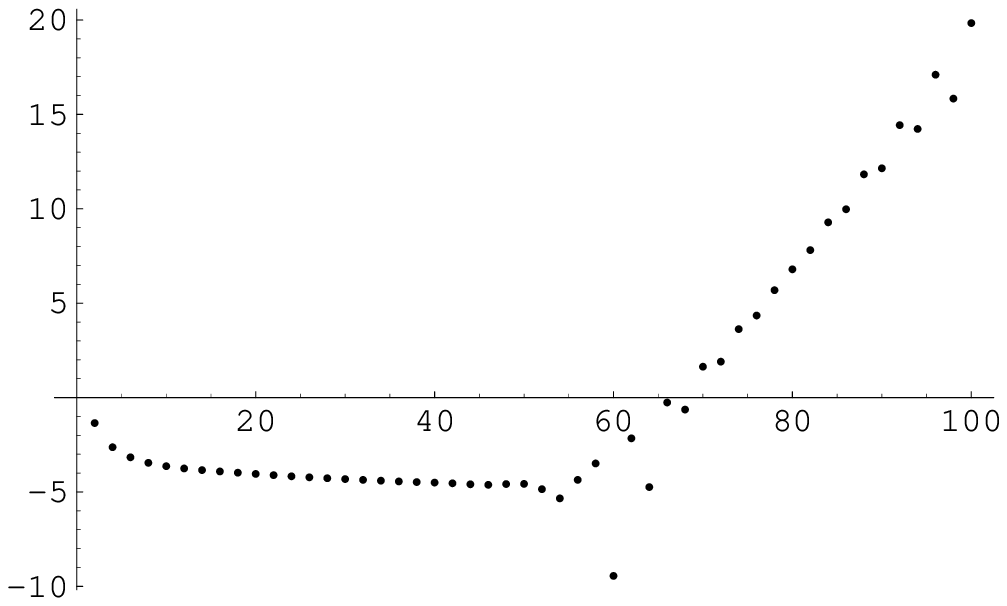}
\caption{\small Plot of $\log |v_{n}|$ for the first 50 even coefficients in the definition
of the wedge state $\ket{\frac 32}$. All the coefficients were calculated exactly as rational
numbers, therefore the irregularity around $n=60$ should be attributed to a chaotic behavior 
rather than to numerical errors.}
}

\newpage

\FIGURE{
\epsfbox{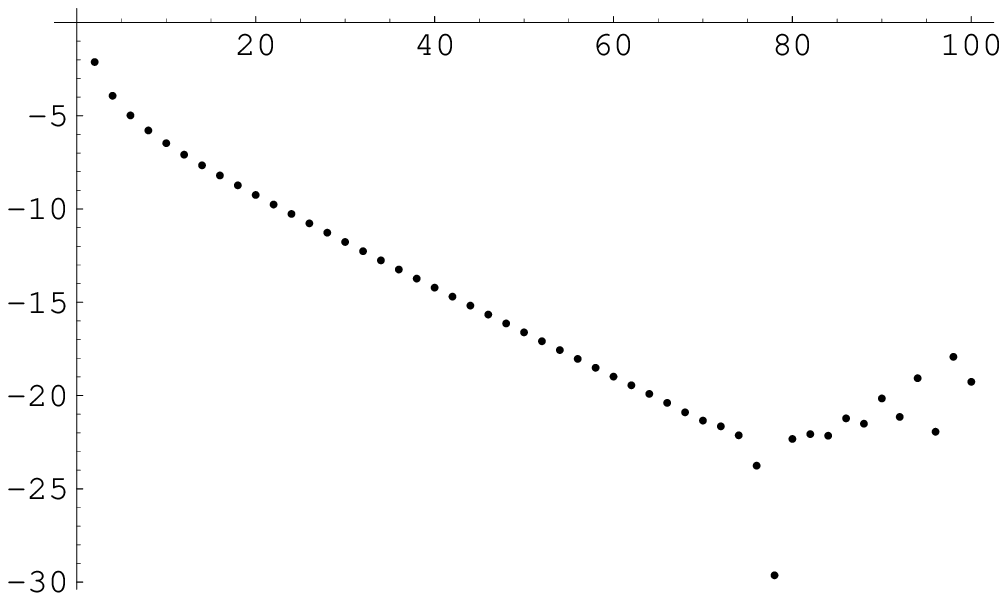}
\caption{\small Plot of $\log |v_{n}|$ for the first 50 even coefficients in the definition
of the wedge state $\ket{\frac 52}$. This state is quite close to the
vacuum, this is the reason why the coefficients decrease exponentially
to a rather high level $n \sim 80$.}
}

\FIGURE{
\epsfbox{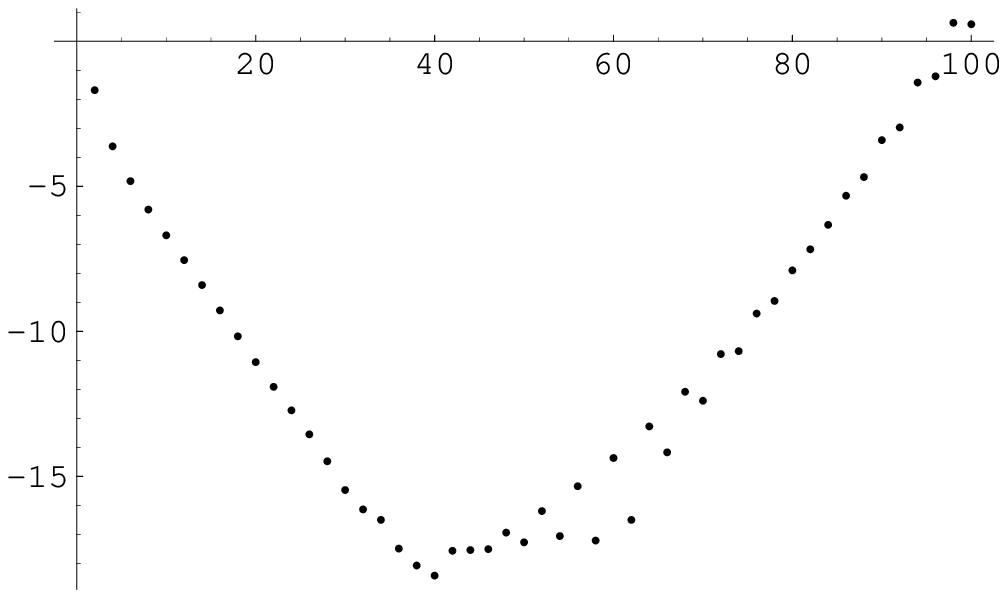}
\caption{\small Plot of $\log |v_{n}|$ for the first 50 even coefficients in the definition
of the wedge state $\ket{3}$.}
}

\newpage

\FIGURE{
\epsfbox{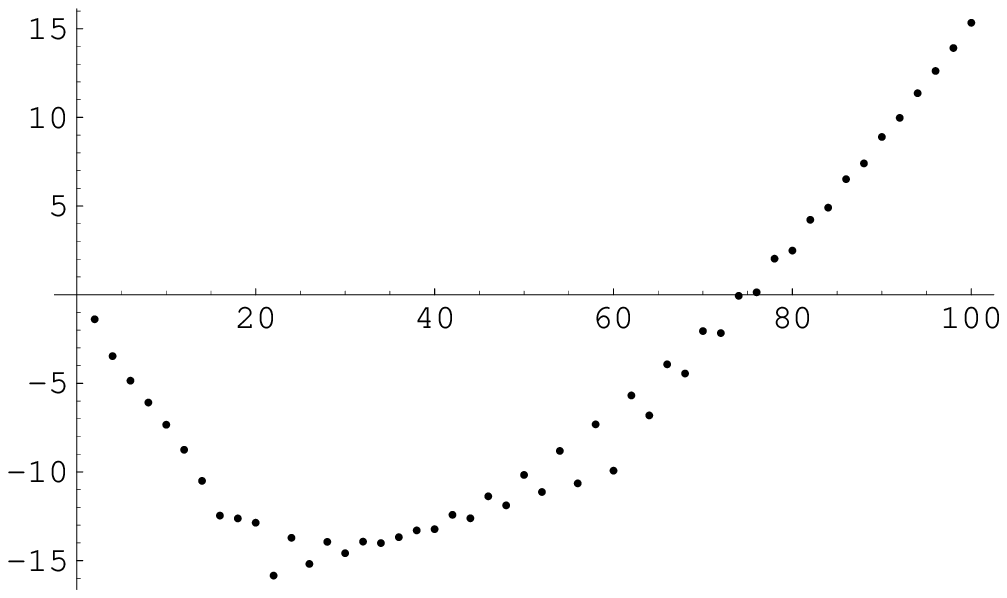}
\caption{\small Plot of $\log |v_{n}|$ for the first 50 even coefficients in the definition
of the wedge state $\ket{4}$.}
}

\FIGURE{
\epsfbox{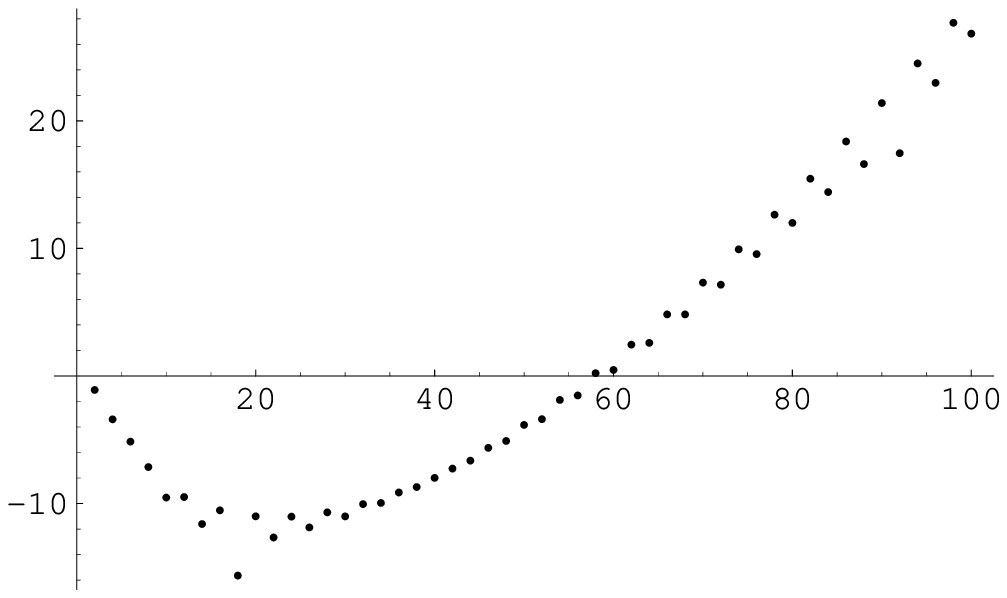}
\caption{\small Plot of $\log |v_{n}|$ for the first 50 even coefficients in the definition
of the sliver state $\ket{\infty}$.}
}

\newpage

\FIGURE{
\epsfbox{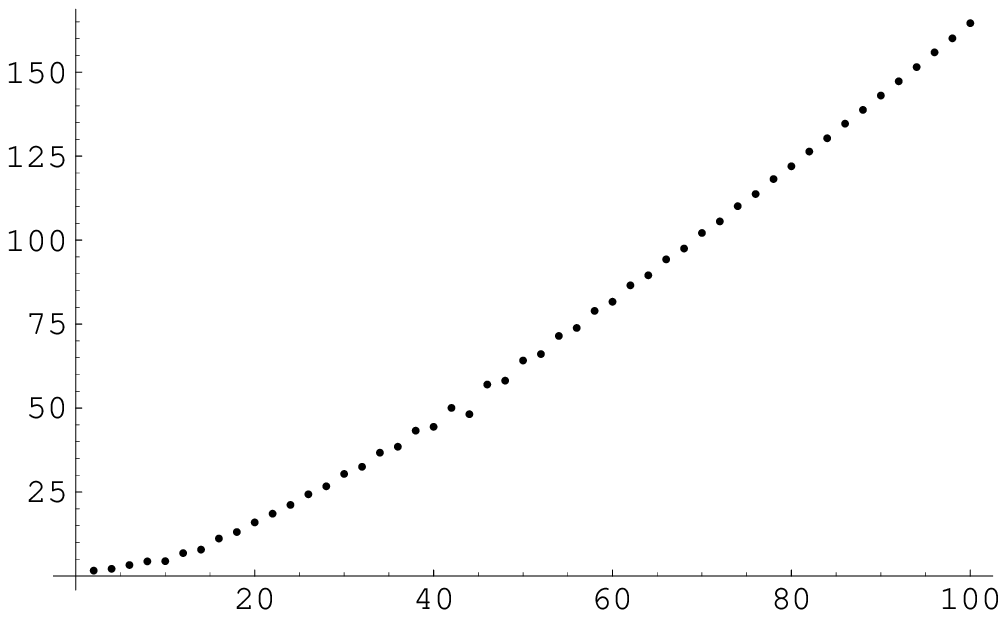}
\caption{\small Plot of $\log |v_{n}|$ for the first 50 even coefficients in the definition
of the wedge state $\ket{\frac12}$. This state is meaningless from the
geometric point of view of conformal field theory. Its star product
with other states in level expansion does not converge as could be
expected from the exponential growth of the coefficients.}
}


\newpage

\begin{thebibliography}{99}

\bibitem{WittenSFT}
E.~Witten,
``Noncommutative geometry and string field theory,''
Nucl.\ Phys.\ B {\bf 268} (1986) 253.

\bibitem{GJ1}
D.~J.~Gross and A.~Jevicki, ``Operator Formulation Of Interacting
String Field Theory I,'' Nucl.\ Phys.\ B {\bf 283} (1987) 1.

\bibitem{GJ2}
D.~J.~Gross and A.~Jevicki, ``Operator Formulation Of Interacting
String Field Theory II,'' Nucl.\ Phys.\ B {\bf 287} (1987) 225.

\bibitem{LPP1}
A.~LeClair, M.~E.~Peskin and C.~R.~Preitschopf, ``String Field
Theory On The Conformal Plane. 1. Kinematical Principles,'' Nucl.\
Phys.\ B {\bf 317} (1989) 411.

\bibitem{LPP2}
A.~LeClair, M.~E.~Peskin and C.~R.~Preitschopf, ``String Field
Theory On The Conformal Plane. 2. Generalized Gluing,'' Nucl.\
Phys.\ B {\bf 317} (1989) 464.

\bibitem{Sen:Descent}
A.~Sen, ``Descent relations among bosonic D-branes,'' Int.\ J.\
Mod.\ Phys.\  A {\bf 14} (1999) 4061 [hep-th/9902105].

\bibitem{Sen:Universality}
A.~Sen, ``Universality of the tachyon potential,'' JHEP {\bf 9912}
(1999) 027 [hep-th/9911116].

\bibitem{SZ}
A.~Sen and B.~Zwiebach, ``Tachyon condensation in string field
theory,'' JHEP {\bf 0003} (2000) 002 [hep-th/9912249].

\bibitem{MT}
N.~Moeller and W.~Taylor, ``Level truncation and the tachyon in
open bosonic string field theory,'' Nucl.\ Phys.\ B {\bf 583}
(2000) 105 [hep-th/0002237].

\bibitem{Ohmori}
K.~Ohmori,
``A review on tachyon condensation in open string field theories,''
hep-th/0102085.

\bibitem{RSZ5}
L.~Rastelli, A.~Sen and B.~Zwiebach,
``Vacuum string field theory,''
hep-th/0106010.

\bibitem{DeSmet}
P.~De Smet,
``Tachyon condensation: Calculations in string field theory,''
PhD thesis, hep-th/0109182.

\bibitem{SchnablPhD}
M.~Schnabl, ``Noncommutative Geometry and String Field Theory'',
PhD thesis, SISSA (2001), unpublished.

\bibitem{Arefeva}
I.~Y.~Arefeva, D.~M.~Belov, A.~A.~Giryavets, A.~S.~Koshelev and P.~B.~Medvedev,
``Noncommutative field theories and (super)string field theories,''
arXiv:hep-th/0111208.

\bibitem{RSZ1}
L.~Rastelli, A.~Sen and B.~Zwiebach,
``String field theory around the tachyon vacuum,''
hep-th/0012251.

\bibitem{GRSZ}
D.~Gaiotto, L.~Rastelli, A.~Sen and B.~Zwiebach,
``Ghost structure and closed strings in vacuum string field theory,''
arXiv:hep-th/0111129.

\bibitem{RSZ4}
L.~Rastelli, A.~Sen and B.~Zwiebach,
``Boundary CFT construction of D-branes in vacuum string field theory,''
hep-th/0105168.

\bibitem{RZ}
L.~Rastelli and B.~Zwiebach, ``Tachyon potentials, star products
and universality,'' hep-th/0006240.

\bibitem{David}
J.~R.~David,
``Excitations on wedge states and on the sliver,''
JHEP {\bf 0107} (2001) 024
[hep-th/0105184].

\bibitem{Samuel}
S.~Samuel, ``The Physical and Ghost Vertices In  Witten's String
Field Theory,'' Phys.\ Lett.\ B {\bf 181} (1986) 255.

\bibitem{Horowitz}
G.~T.~Horowitz and A.~Strominger,
``Translations As Inner Derivations And Associativity Anomalies In Open String Field Theory,''
Phys.\ Lett.\ B {\bf 185} (1987) 45.

\bibitem{QS}
Z.~Qiu and A.~Strominger,
``Gauge Symmetries In (Super)String Field Theory,''
Phys.\ Rev.\ D {\bf 36} (1987) 1794.

\bibitem{HM}
G.~T.~Horowitz and S.~P.~Martin,
``Conformal Field Theory And The Symmetries Of String Field Theory,''
Nucl.\ Phys.\ B {\bf 296} (1988) 220.

\bibitem{Romans}
L.~J.~Romans, ``Operator Approach To Purely Cubic String Field
Theory,'' Nucl.\ Phys.\ B {\bf 298} (1988) 369.

\bibitem{EFHM}
I.~Ellwood, B.~Feng, Y.~He and N.~Moeller,
``The identity string field and the tachyon vacuum,''
JHEP {\bf 0107} (2001) 016
[hep-th/0105024].

\bibitem{Matsuo2}
Y.~Matsuo,
``Identity projector and D-brane in string field theory,''
Phys.\ Lett.\ B {\bf 514} (2001) 407
[arXiv:hep-th/0106027].

\bibitem{TT}
T.~Takahashi and S.~Tanimoto,
``Wilson lines and classical solutions in cubic open string field theory,''
Prog.\ Theor.\ Phys.\  {\bf 106} (2001) 863
[arXiv:hep-th/0107046].

\bibitem{Kishimoto}
I.~Kishimoto,
``Some properties of string field algebra,''
JHEP {\bf 0112} (2001) 007
[arXiv:hep-th/0110124].

\bibitem{KO}
I.~Kishimoto and K.~Ohmori,
``CFT description of identity string field: Toward derivation of the VSFT  action,''
arXiv:hep-th/0112169.

\bibitem{KP}
V.~A.~Kosteleck\'y and R.~Potting,
``Analytical construction of a nonperturbative vacuum for the open  bosonic string,''
Phys.\ Rev.\ D {\bf 63} (2001) 046007
[arXiv:hep-th/0008252].

\bibitem{RSZ2}
L.~Rastelli, A.~Sen and B.~Zwiebach,
``Classical solutions in string field theory around the tachyon vacuum,''
hep-th/0102112.

\bibitem{RSZ3}
L.~Rastelli, A.~Sen and B.~Zwiebach,
``Half strings, projectors, and multiple D-branes in vacuum string field  theory,''
JHEP {\bf 0111} (2001) 035
[arXiv:hep-th/0105058].

\bibitem{GT}
D.~J.~Gross and W.~Taylor,
``Split string field theory. I,''
JHEP {\bf 0108} (2001) 009
[arXiv:hep-th/0105059].

\bibitem{Matsuo1}
Y.~Matsuo,
``BCFT and sliver state,''
Phys.\ Lett.\ B {\bf 513} (2001) 195
[arXiv:hep-th/0105175].

\bibitem{Mukhopadhyay}
P.~Mukhopadhyay,
``Oscillator representation of the BCFT construction of D-branes in  vacuum string field theory,''
JHEP {\bf 0112} (2001) 025
[arXiv:hep-th/0110136].

\bibitem{Moeller}
N.~Moeller,
``Some exact results on the matter star-product in the half-string  formalism,''
arXiv:hep-th/0110204.

\bibitem{Moore}
G.~Moore and W.~Taylor, ``The singular geometry of the sliver,''
arXiv:hep-th/0111069.

\bibitem{AGM}
I.~Y.~Aref'eva, A.~A.~Giryavets and P.~B.~Medvedev,
``NS matter sliver,''
arXiv:hep-th/0112214.

\bibitem{MS}
M.~Marino and R.~Schiappa,
``Towards vacuum superstring field theory: The supersliver,''
arXiv:hep-th/0112231.

\bibitem{Bonora}
L.~Bonora, D.~Mamone and M.~Salizzoni,
``B field and squeezed states in vacuum string field theory,''
arXiv:hep-th/0201060.

\bibitem{Kuczma}
M.~Kuczma, B.~Choczewski and R.~Ger,
``Iterative Functional Equations'' Cambridge University Press, 1990

\bibitem{RSZ7}
L.~Rastelli, A.~Sen and B.~Zwiebach, ``Star algebra
spectroscopy,'' arXiv:hep-th/0111281.

\bibitem{RT}
M.~Rakowski and G.~Thompson,
``On The Associativity Anomaly In Open String Field Theory,''
Phys.\ Lett.\ B {\bf 197} (1987) 339.

\bibitem{Schnabl2}
M.~Schnabl,
``Constraints on the tachyon condensate from anomalous symmetries,''
Phys.\ Lett.\ B {\bf 504} (2001) 61
[hep-th/0011238].

\bibitem{Schnabl4}
M.~Schnabl,
``Anomalous reparametrizations and butterfly states in string field  theory,''
arXiv:hep-th/0202139.


\end{thebibliography}
\end{document}